\documentclass[twocolumn,times]{aastex63}

\usepackage[caption=false]{subfig}

\received{April 20, 2021}
\revised{May 5, 2021}
\accepted{May 7, 2021}

\shorttitle{RR Lyrae binaries I.}
\shortauthors{Hajdu et al.}

\begin{document}

\title{Studies of RR~Lyrae Variables in Binary Systems. I.: Evidence of
a Trimodal Companion Mass Distribution}

\correspondingauthor{Gergely Hajdu}
\email{ghajdu@camk.edu.pl}

\author[0000-0003-0594-9138]{Gergely Hajdu}
\affiliation{Nicolaus Copernicus Astronomical Center, Polish Academy of Sciences,
Bartycka 18,
Warsaw, 00-716, Poland}

\author[0000-0002-9443-4138]{Grzegorz Pietrzy\'nski}
\affiliation{Nicolaus Copernicus Astronomical Center, Polish Academy of Sciences,
Bartycka 18,
Warsaw, 00-716, Poland}

\author[0000-0001-7901-7689]{Johanna Jurcsik}
\affiliation{Konkoly Observatory, Research Centre for Astronomy and Earth Sciences, E\"{o}tv\"{o}s Lor\'{a}nd Research Network (ELKH),
Konkoly Thege Mikl\'os \'ut 15-17.,
Budapest, H-1121, Hungary}

\author[0000-0001-7901-7689]{M\'arcio Catelan}
\affiliation{Pontificia Universidad Cat\'olica de Chile, Facultad de F\'isica, Instituto de Astrof\'isica,
Av. Vicu\~na Mackenna 4860,
7820436 Macul, Santiago, Chile}
\affiliation{Millennium Institute of Astrophysics,
Nuncio Monse\~{n}or Sotero Sanz 100, Of. 104,
Providencia, Santiago, Chile}

\author[0000-0002-0136-0046]{Paulina Karczmarek}
\affiliation{Universidad de Concepci\'on, Departamento de Astronom\'ia,
Casilla 160-C,
Concepci\'on, Chile}

\author[0000-0003-3861-8124]{Bogumi\l{} Pilecki}
\affiliation{Nicolaus Copernicus Astronomical Center, Polish Academy of Sciences,
Bartycka 18,
Warsaw, 00-716, Poland}

\author[0000-0002-7777-0842]{Igor Soszy\'nski}
\affiliation{Astronomical Observatory, University of Warsaw,
Aleje Ujazdowskie 4,
Warsaw, 00-478, Poland}

\author[0000-0001-5207-5619]{Andrzej Udalski}
\affiliation{Astronomical Observatory, University of Warsaw,
Aleje Ujazdowskie 4,
Warsaw, 00-478, Poland}

\author{Ian B. Thompson}
\affiliation{The Observatories of the Carnegie Institution for Science,
813 Santa Barbara Street,
Pasadena, CA 91101, USA}

\begin{abstract}

We present 87 candidates for RR~Lyrae variable stars in binary systems,
based on our new search using the light-travel time effect (LTTE) and observed -- calculated ($O-C$)
diagrams in the Galactic bulge time-series photometry of the Optical Gravitational Lensing Experiment.
Out of these, 61 are new candidates, while 26 have been announced
previously. Furthermore, 12 stars considered
as binary candidates in earlier works are discarded from the list, either because they were found to
have $O-C$ diagrams incompatible with the LTTE or because
their long-term periodicity is definitely caused by the Blazhko effect.

This sample of RR~Lyrae binary candidates allows us to draw the first firm conclusions about
the population of such objects: no candidate has an orbital period below 1000\,days, while their
occurrence rate steadily increases with increasing period, and peaks between
3000 and 4000\,days; however, the decrease in the number of stars toward even longer periods
is probably the result of observational biases. The eccentricities show a very
significant concentration between 0.25 and 0.3, with a quarter of candidates found in this single bin,
overlaid on an otherwise flat distribution between 0.05 and 0.6. Only six stars have higher
inferred eccentricities above 0.6.
Lastly, the distribution of the mass functions is highly peculiar, exhibiting strong trimodality.
We interpret these modes as the presence of three distinct groups of companions, with typical
inferred masses of $\sim0.6$, $\sim0.2$, and $\sim0.067\,\mathrm{M}_\odot$, which can be associated with
populations of white dwarf and main sequence, red dwarf, and brown dwarf companions, respectively.

\end{abstract}


\section{Introduction} \label{sec:intro}

After more than 100 years of continuous study, RR~Lyrae variables (hereafter RRLs)
remain at the forefront of astrophysics, including their use as
distance indicators toward old stellar populations (e.g., \citealt{2017AJ....154..263K,2018SSRv..214..113B}), the study of the 3D structure
of old stellar populations in nearby galaxies (e.g., \citealt{2017AcA....67....1J,2018ApJ...859...31H,2020MNRAS.495.4124F}),
the behavior of interstellar extinction (e.g., \citealt{2019ApJ...874...30S}),
and even as probes of their metallicity distributions (e.g., \citealt{2016MNRAS.461L..41M,2020MNRAS.492.1061V}).

Despite our extensive knowledge of these variables, and the large number of members
(well over $100{,}000$) discovered in the Milky Way and its satellite galaxies by sky surveys such as
the Optical Gravitational Lensing Experiment
(OGLE; Magellanic System: \citealt{2016AcA....66..131S}; Milky Way bulge and disk: \citealt{2019AcA....69..321S}),
the variability survey of the \textit{Gaia} satellite \citep{2019A&A...622A..60C},
the Catalina Sky Survey \citep{2015MNRAS.446.2251T}, Pan-STARRS \citep{2017AJ....153..204S},
and the Vista Variables in the V\'ia L\'actea \citep{2020ApJ...898...46D} among others,
the arguably most important stellar property, the stellar mass,
has never been measured directly even for a single member of this class.
As the most straightforward way to measure stellar masses is by characterizing
the orbits of members of binary systems, searches for binary RRLs
have been carried out, but until very recently, only the variable TU~UMa had been
demonstrated to reside in one with high certainty
\citep{1999AJ....118.2442W,2016A&A...589A..94L}. The very long ($\sim23$\,yr) orbital period,
combined with no detection of light from the companion, hinders the complete characterization
of this system.

Sky surveys have found very few candidates for RRLs in eclipsing binary systems,
with spectroscopic follow-up confirming them as either blends \citep{2008A&A...489.1209P}
or, in the case of OGLE-BLG-RRLYR-02792, an entirely new kind of variable star, a so-called
binary evolution pulsator (BEP; \citealt{2012Natur.484...75P,2013MNRAS.428.3034S}).
This is not an entirely unexpected result: in a close binary system Roche-lobe
overflow would happen while the star traverses the red giant branch. This would
result in too much mass loss before it evolves onto the horizontal branch (HB)
to settle inside the main instability strip, resulting in a blue HB star instead
(see, e.g., \citealt{2020A&A...641A.163V}, as well as references therein).
Therefore, bona fide binary RRLs are necessarily located in wide binary systems
(with orbital periods longer than around 1000\,days), greatly diminishing the
probability of finding one in an eclipsing system.

The most widely utilized method in the search for RRL binaries uses the
light-travel time effect (LTTE; \citealt{1952ApJ...116..211I,1959AJ.....64..149I}), the modulation
of the light curves induced by the changing distance between the observer and the
pulsating component in a binary system due to its motion around the center of mass.
Most commonly, these changes are quantified using the so-called observed -- calculated ($O-C$) diagrams
\citep{2005ASPC..335....3S}, which are constructed using the timings of specific parts of the light curve of the variable in question
(e.g., maxima of the light curve, or the middle of the rising branch for pulsating variables; light-curve
minima for eclipsing variables) and the period of the variable. This diagram is then analyzed looking for the characteristic shape expected
of the LTTE. This method was used in the context of the high-precision Kepler space telescope observations
\citep{2014MNRAS.444..600L,2015EPJWC.10106030G,2017MNRAS.465L...1S}, as well as for field RRL variables that can have $O-C$ diagrams
spanning over 100\,years \citep{2004MNRAS.354..821D,2016MNRAS.459.4360L,2018PASJ...70...71L,2018ApJ...863..151L} to identify a
number of binary candidates. However, it should be noted that in the only published case where radial velocity observations were
available to revise the claimed binarity, this possibility has been conclusively discarded (Z~CVn; \citealt{2018MNRAS.474..824S}).

An alternative way to construct $O-C$ diagrams is by following the proposal of \citet{1919AN....210...17H}, and measuring
the phase shift of the whole light curve using a template. This technique is particularly well suited for use on data
obtained by wide-field variability surveys, where the random nature of the time sampling usually results in well-covered light curves.
For such data sets, measuring the times of individual maxima may actually be impossible, depending on the sparsity of the
sampling. However, a significant drawback of this method is that care must be taken both during the construction of the
light-curve template and during the definition of the sections of the light curves, because choices made for these crucial points
can have a drastic effect on the quality of the $O-C$ diagram.
Nevertheless, we have utilized this method in our initial search for RRL binary candidates \citep{2015MNRAS.449L.113H} with the
OGLE survey light curves \citep{2015AcA....65....1U} of RRLs toward the Galactic bulge.
The aforementioned drawbacks were reduced mostly by resorting to an iterative process for the construction of the
template light curves, where we have corrected the light curves for an initial binary $O-C$ solution, and refitted the template
to these corrected data. These new templates, which resembled the intrinsic light-curve shape of their respective stars
much better (when compared to the initial ones), allowed the derivation of much more accurate $O-C$ points.
The binary RRL candidate sample was further extended by \citet{2019MNRAS.487L...1P}, who used similar procedures to those described in \citet{2015MNRAS.449L.113H}.

It is worth mentioning that traditional $O-C$ diagrams are not the only way of characterizing the LTTE:
phase-modulated light curves in Fourier space exhibit characteristic structures that
can also be modeled, allowing the determination of orbital parameters with high accuracy \citep{2012MNRAS.422..738S,2015MNRAS.450.3999S}.
This technique has been tried for RRL stars already \citep{2017EPJWC.15203006S}, although it
might not be entirely suited for the study of this particular class of variables: generally the available data span only a few times the
length of the supposed orbital periods, resulting in wide peaks for the characteristic frequencies in Fourier space.
This, as well as the intrinsic, slow change of the period of pulsation, limits the accuracy of the technique.
Nevertheless, it is a very valuable tool for
shorter-period variables (e.g., $\delta$~Scuti, SX~Phe, or various classes of pulsating white dwarfs),
allowing us to study their general binarity parameters \citep{2018MNRAS.474.4322M}, and sensitive enough to
discover exoplanets \citep{2016ApJ...827L..17M}.
Another method, forward modeling of modulated light curves, has been proposed as an alternative characterization of the LTTE
very recently \citep{2020AJ....159..202H}. This technique holds a lot of promise, and its applicability should be explored
in the context of RRL stars in the future. 

Regardless of the exact method, the information learned from the LTTE for a well characterized
variable is the same as that provided by radial velocity observations of a single-lined spectroscopic binary (SB1):
the projected orbit of the variable is characterized, but no information is provided on either
of the inclination of the system or on the mass ratio of its components. For this reason, additional
constraints (radial velocity follow-up; astrometry; theoretical considerations) are a must for
learning more about these systems.

Besides the LTTE, radial velocity observations can also be used to discover RRLs in binary systems, similarly
to other types of stars. However, this is significantly more difficult for RRL variables, because their
rapid, large-amplitude pulsations must be disentangled from the expected long-term, small-amplitude binary signal.
There is one project currently attempting this \citep{2016CoKon.105..145G}; however, it has not yet resulted
in the announcement of new RRL binary candidates.

The detection of binary RRLs through their astrometric motion
\citep{2019A&A...623A.116K}, as well as identification of their resolved companions by common proper
motions \citep{2019A&A...623A.117K} and high-resolution imaging techniques \citep{2020RNAAS...4..143S},
has been attempted only very recently. While these techniques are very promising, full characterization
of a binary orbit might still be unattainable if the companion is very faint (e.g., a stellar remnant)
or if the orbital period is too long for follow-up observations.

Finally, RRL variables with chemical abundance anomalies, such as carbon enhancement
\citep{2011A&A...527A..65H,2014ApJ...787....6K} and even overabundance of neutron-capture elements (TY~Gru; \citealt{2006AJ....132.1714P})
might be explained by pollution of the stellar atmosphere by earlier mass transfer from an asymptotic giant branch (AGB)
companion \citep{2013MNRAS.435..698S}. Nevertheless, none of these stars have been confirmed to reside
in a binary system so far.

In this new series of papers, we are undertaking a systematic study of binary
RRL variables, and in this initial paper we are specifically searching for new
candidates in the OGLE Galactic bulge light curves. As some of these data
have been searched before both by us \citep{2015MNRAS.449L.113H,2018pas6.conf..248H}
and by an independent group \citep{2019MNRAS.487L...1P}, we also revise the list
of prospective RRL binaries announced so far. Our final candidate list now has enough variables
to draw meaningful conclusions about the companion population of RRLs in binary systems, even
if some of them will likely turn out to be false candidates in the future.

\section{Data and the search for binary candidates} \label{sec:search}

During this new search, we have used an updated catalog of the OGLE RRL variables toward the Galactic bulge, published
by \citet{2014AcA....64..177S}, with the $I$-band  light curves including observations until the 2017 observing season.
All OGLE observations were obtained with the 1.3\,m Warsaw telescope at Las Campanas observatory \citep{2015AcA....65....1U}.
Although many more RRLs have been published by the OGLE project in the Galactic bulge and disk owing to the OGLE Galaxy
Variability Survey \citep{2019AcA....69..321S}, their light curves are generally too short to conduct a search for
binary signals that are multiple years long.
As in our previous work \citep{2015MNRAS.449L.113H}, here we consider only fundamental-mode RRL
variables (RRab subtype).

\subsection{Construction of the initial $O-C$ diagrams} \label{subsec:initial}

Our initial $O-C$ diagrams were constructed using similar, but slightly modified procedures to those in \citet{2015MNRAS.449L.113H},
expanding upon the original method of \citet{1919AN....210...17H}.
It is important to note that stars observed during both the OGLE-III and IV phases of the OGLE project
have effectively two independent, non-overlapping $I$-band light curves. Although these light curves were derived using similar procedures,
one must exercise caution when attempting to combine them, because they use different cameras and filters (see \citealt{2015AcA....65....1U} for an overview),
therefore can have different magnitude zero-points.
The OGLE project uses image subtraction combined with zero points derived using
point-spread function photometry. While image subtraction is expected to produce consistent results on the \textit{intensity scale},
even a small error in the zero point for either the OGLE-III or IV (or both) part(s) of the light curve will result in
different average magnitudes. Furthermore, the brighter of the two parts will have diminished pulsation amplitude on the \textit{magnitude scale}
when compared to the fainter section.
Therefore, as a first step, for each star we have transformed both the OGLE-III and IV photometries to an arbitrary intensity scale.
Both sections of the light curve were then fitted independently with a fifth-order truncated Fourier series, and the OGLE-III photometry was shifted
to the mean level of the OGLE-IV observations, before finally transforming it back to the magnitude scale.
We note that in \citet{2015MNRAS.449L.113H} the difference between the OGLE-III and IV
photometry was only corrected for by a simple shift in magnitude.

During the second step, the homogenized (if both OGLE-III and OGLE-IV data were available) light curves were folded using their
periods given by \citet{2014AcA....64..177S}, and fitted with a fifth-order truncated Fourier series using ordinary least-squares (OLS)
regression to serve as the light-curve template. Then, the light curve
was divided first into observing seasons, and each season was further divided into two segments if more than 40 light-curve points were
available (in \citealt{2015MNRAS.449L.113H}, this was done if within a season the baseline of observations exceeded 160\,days,
but sometimes this resulted in sections having very few points, hence the change).
The $O-C$ points were then derived as the phase shift that minimized the scatter between the light-curve template and the points
in each of the individual segments.

In total, $27{,}480$ $O-C$ diagrams were constructed in this way for RRab variables (with the last one being OGLE-BLG-RRLYR-39119; hereafter,
every OGLE bulge RRL star will be referred to only by its numerical ID), and all of them were inspected, simultaneously with
their light curves (both original and folded). The average and median numbers of data points were 2483 and 766, while the average
and median time baselines were 10.5 and 7.4 yr, respectively.
As mentioned in Section~\ref{sec:intro}, the exact way in which the template light curve
is constructed can have a big impact on the individual $O-C$ measurements, possibly introducing biases that could hinder
the recognition of the LTTE. In order to compensate for this, we have selected all candidates with $O-C$ shapes even vaguely
resembling the expected LTTE for further analysis, resulting in a list of $\sim400$ RRL variables.
Contrary to the restrictive approach employed in \citet{2015MNRAS.449L.113H},
we have not discarded stars showing the Blazhko effect (\citealt{1907AN....175..325B}; see also \citealt{2016pas..conf...22S}
for a recent review and additional references), and also included all previously announced binary candidates from
\citet{2015MNRAS.449L.113H,2018pas6.conf..248H} and \citet{2019MNRAS.487L...1P}, as well.

\subsection{Analysis of the Light-travel Time Effect} \label{subsec:bin}

The effect of the LTTE on the $O-C$ diagrams can be modeled by its characteristic shape, following \citet{1952ApJ...116..211I,1959AJ.....64..149I}.
Throughout this work, we have adopted the LTTE shape as

\begin{equation}
    \label{eq:lte}
    z(t) = a_1\sin i \frac{1-e^2}{1+e \cos(\nu)} \sin (\nu + \omega),
\end{equation}

\noindent where $z(t)$ is the time-dependent distance between the observer and the variable component of the binary,
$a_1 \sin i$ is the projected semimajor axis of the RRL, $e$ is the eccentricity, $\omega$ is the argument of the periastron,
and $\nu$ is the true anomaly. We recall that the latter quantity can be calculated for any point in time through

\begin{equation}
    \label{eq:true_anomaly}
    \cos \nu = \frac{\cos E - e}{1-e \cos E},
\end{equation}

\noindent where the eccentric anomaly $E$ is connected to the mean anomaly $M$ by

\begin{eqnarray}
    \label{eq:eccentric_anomaly}
    M = E - e\sin E = \frac{2\pi}{P_\mathrm{orb}} \left(t-T_0\right),
\end{eqnarray}

\noindent which itself is a linear function of time, depending on the orbital period $P_\mathrm{orb}$,
the time of periastron passage $T_0$, and the time of observations $t$.

Unfortunately, the LTTE interpretation of the $O-C$ diagrams is complicated by the fact that changes in the variability
period accumulate over time. When this change is approximately linear, as is expected from evolutionary changes
on human time scales for RRL stars, this causes an additional, parabolic shape to be present in the $O-C$ diagram. We can
extend Equation~\ref{eq:lte} to model it as

\begin{eqnarray}
    \label{eq:parabola}
    (O-C)(t) = z(t) + c_0 + c_1 \cdot t + c_2 \cdot t^2,
\end{eqnarray}

\noindent where $c_1$ is the fractional difference between the period used to construct the $O-C$ diagram and
the pulsational period at $t=0$, while $c_2$ is related to the linear period-change rate \citep{2005ASPC..335....3S} as:

\begin{eqnarray}
    \label{eq:pchange}
    \beta = c_2 \cdot 2 \cdot P,
\end{eqnarray}

\noindent where $P$ is the pulsational period of the RRL in question adopted for the construction of its $O-C$ diagram.\footnote{
It should be noted that this formula is different from that presented by \citet{2005ASPC..335....3S},
where the variability period is in the denominator. The reason for this is the choice of the independent variable:
in our case it is time $t$, while \citet{2005ASPC..335....3S} considers the epoch number $E$ (not to be confused with the eccentric
anomaly in Eqs.~\ref{eq:true_anomaly} and \ref{eq:eccentric_anomaly}) as the independent variable.
}
The period-change rate $\beta$ is usually expressed in units of $\mathrm{day}\,\mathrm{Myr}^{-1}$ for RRL
variables. The (absolute) value of this parameter is generally smaller than $0.5\,\mathrm{day}\,\mathrm{Myr}^{-1}$
for most regular RRLs (see, e.g., \citealt{2001AJ....121..951J,2011MNRAS.411.1744S,2012MNRAS.419.2173J}).

For the initial analysis of the $\sim400$ stars selected in Section~\ref{subsec:initial}, we used
a very similar iterative procedure to that outlined in \citet{2015MNRAS.449L.113H}, but also
supplemented the light curves with OGLE observations taken in 2018 and
2019.\footnote{The extended light curves will be part of newer editions of the OGLE Catalog of Variable
Stars. In the meantime they can be acquired from the OGLE team upon request.}
The initial $O-C$ diagrams were prepared for this data set as described in Section~\ref{subsec:initial}.
Then, the $O-C$ points were fit with Equation~\ref{eq:parabola} to
model the shape of the $O-C$ diagram expected from the combination of the LTTE and a linear period change of
the pulsational period of the RRL variables. The fit was then subtracted from the timings of the individual
$I$-band observations, resulting in modified light curves mostly devoid of the effect of both the LTTE and
the linear period change.

The modified light curve was used to construct an improved template, fit with a higher-order 
Fourier series, which follows the real light-curve shape much better, resulting in improved $O-C$ estimates
when using them to redetermine their values.
Furthermore, the uncertainties of these new $O-C$ points were also estimated using bootstrapping,
by drawing 500 new samples for each of the light-curve segments (for $N$ points in a segment we drew $N$ points
with replacement). For each of the segments, the $O-C$ values of these 500 samples were also determined,
and their standard deviations with respect to the non-bootstrapped estimates were adopted as their uncertainties.

This procedure was repeated multiple times for each of the variables, while simultaneously varying the
assumed pulsational period and the orders of the Fourier harmonics applied in each step of the iterations.
At this point, stars were removed from the list for a variety of reasons. Some of these variables displayed clear
(albeit sometimes tiny, $\sim0.01$\,mag) amplitude modulation with the same period as the supposed binary motion.
Other stars deviated from the periodic $O-C$ behavior when the light curve was extended with the last two seasons
of OGLE observations. In the case of many variables, mostly those with long purported binary periods and low amplitudes,
it could not be decided from the available data whether a binary model was more favorable than a nonlinear period-change
behavior. Therefore, we have discarded them from further analysis. All in all, fewer than 100 variables remained after this step in our analysis.

\subsection{Determination of the final parameters with MCMC} \label{subsec:mcmc}

We continued analyzing the remaining candidates by further modifying the iterative procedure outlined in Section~\ref{subsec:bin},
changing it at three key points.

First, due to the short binary periods of some of the binary candidates, it became necessary to divide the light curves into
smaller parts, because the expected $O-C$ values within a season could change more than their formal errors. Therefore, we
we adopted the following (default) scheme for dividing seasons into smaller segments: seasons with fewer than 9 points were not
analyzed; those with $9-80$ points were left as one segment; seasons with $81-120$, $121-160$, $161-240$, and $241$
or more points points were divided into 2, 3, 4, and 5 segments, respectively.
This division was only modified only in special cases (see Section~\ref{sec:parameters}).

Second, it was recognized that in cases of eccentricities with large errors (usually when the eccentricity itself is small or
the $O-C$ diagram is noisy), the behavior of the argument of periastron can become erratic. This can be thought of as a result of
the latter quantity losing its meaning at small eccentricities: when $e=0$, the orbit is circular, and all points along it
can be regarded as the periastron. Therefore, during the second iteration of the $O-C$ fitting procedure, instead of fitting
the values of $e$ and $\omega$ directly, we consider
them as polar coordinates, and use their Cartesian equivalents $\sqrt{e}\sin\omega$ and $\sqrt{e}\cos\omega$
as free parameters.
Then, the values of the original quantities can be recovered trivially as follows:

\begin{eqnarray}
    \label{eq:econv}
    e' &=& (\sqrt{e}\sin\omega)^2 + (\sqrt{e}\cos\omega)^2, \\
    \label{eq:omegaconv}
    \omega' &=& \mathrm{arctan2} (\sqrt{e}\sin\omega,\sqrt{e}\cos\omega),
\end{eqnarray}

\noindent which is similar to the conversion between parameters of the linear and nonlinear forms of harmonics of a Fourier series.

Finally, in the course of the second iteration of the procedure, for the estimation of the final binary parameters and their uncertainties,
we have adopted a Bayesian formalism, leveraging the ensemble Markov Chain Monte Carlo (MCMC) capabilities of the
\textsc{emcee} package \citep{2013PASP..125..306F}. The calculations were done for all stars using the default sampler, implementing the
affine-invariant ensemble stretch move presented by \citet{2010CAMCS...5...65G}. The current implementation of \textsc{emcee} supports
alternative samplers, and even allows the use of a weighted mix of samplers during the MCMC procedure. We have conducted limited testing
of these possibilities, but have found that in the case of our specific problem they generally have very similar or clearly inferior performance
(for example, in terms of convergence and autocorrelation lengths) to the default one, and therefore we decided not to use them.

As our routines implementing the calculation of the $O-C$ shape (Eqs.~\ref{eq:lte} and \ref{eq:parabola}) are somewhat
inefficient due to the usage of the \textsc{Python} programming language, we have experimented with the \textsc{Numba} \citep{Numba}
package to speed up critical parts of the computations. Using \textsc{Numba} with minimal changes to the code, 
a speedup of $\times 3-5$ was achieved during the MCMC fitting, allowing us to run longer chains and experiment with the fitting process more.
In order to aid with the reproducibility of our results, the subroutines implementing the method described
in this section are available on GitHub\footnote{\url{https://github.com/gerhajdu/rrl_binaries_1}}, along with supporting Jupyter notebooks \citep{jupyter}
showcasing their use for RRL stars.

\section{The adopted LTTE solutions of the binary RR~Lyrae candidates} \label{sec:parameters}

We have used the iterative $O-C$ procedure, with the MCMC implementation in the second iteration outlined in Section~\ref{subsec:mcmc}, to
continue analyzing the stars remaining after the initial examination of prospective RRL binary candidates (Sect.~\ref{subsec:initial}).
We choose to alter the process of dividing seasons into smaller segments for five variables: for 06992 and 08830 the data from 2001 were
not used; for 09698 and 10142 the data from 2001 and 2002 were merged into one $O-C$ point to stabilize the LTTE solutions.
Furthermore, due to the quick changes of the $O-C$ values (the combined result of a large period-change rate and a relatively short-period,
large-amplitude LTTE component) for 20376, we decided to split all seasons into two sections, even though not all of them had more than 80 points.

\startlongtable
\begin{deluxetable*}{cc|lrrrlr|rcc|c}
    \tablecaption{Parameters of the binary RR~Lyrae sample from the MCMC analysis \label{tab:binprop}}
    \tablewidth{700pt}
    \tabletypesize{\scriptsize}
    \tablehead{
    \colhead{ID} & \colhead{$P$} & 
    \colhead{$P_\mathrm{orb}$} & \colhead{$a_1 \sin i$} &  \colhead{$e$} & \colhead{$\omega$} & \colhead{$T_0$} & \colhead{$\beta$} &
    \colhead{$K_1$} & \colhead{$f(m)$} & \colhead{$M_{\mathrm{S,min.}}$} & \colhead{Quality}\\ 
    \colhead{} & \colhead{(days)} & 
    \colhead{(days)} & \colhead{(au)} & \colhead{} & \colhead{(deg)} & \colhead{(days)} & \colhead{(day\,Myr$^{-1}$)} &
    \colhead{(km s$^{-1}$)} & \colhead{($M_\odot$)} & \colhead{($M_\odot$)}
    } 
    \startdata
    02387 & 0.6722870 & 1947 $\pm$  14 & 1.562 $\pm$ 0.084 & 0.370 $\pm$ 0.092 &  121 $\pm$ 16 & 8738 $\pm$  87 &  0.070 $\pm$  0.044 &  9.48 $\pm$ 0.85 & 0.13521 $\pm$ 0.02265 & 0.5935 & Q1\\
    02854 & 0.6002180 & 3478 $\pm$  20 & 2.328 $\pm$ 0.079 & 0.454 $\pm$ 0.052 &  128 $\pm$  6 & 8250 $\pm$  59 & -0.061 $\pm$  0.027 &  8.20 $\pm$ 0.45 & 0.13961 $\pm$ 0.01428 & 0.6029 & Q1\\
    02950 & 0.4951840 & 2572 $\pm$   8 & 2.321 $\pm$ 0.041 & 0.295 $\pm$ 0.029 & -159 $\pm$  5 & 6423 $\pm$  38 &  0.075 $\pm$  0.017 & 10.28 $\pm$ 0.26 & 0.25230 $\pm$ 0.01368 & 0.8151 & Q1\\
    04376 & 0.4910180 & 2909 $\pm$  19 & 0.755 $\pm$ 0.022 & 0.423 $\pm$ 0.059 & -104 $\pm$  7 & 6505 $\pm$  54 &  0.071 $\pm$  0.005 &  3.13 $\pm$ 0.18 & 0.00680 $\pm$ 0.00064 & 0.1653 & Q2\\
    04628 & 0.5850070 & 2164 $\pm$  30 & 0.696 $\pm$ 0.035 & 0.270 $\pm$ 0.082 &  177 $\pm$ 15 & 7279 $\pm$  91 &  0.064 $\pm$  0.024 &  3.66 $\pm$ 0.28 & 0.00971 $\pm$ 0.00161 & 0.1899 & Q2\\
    04837 & 0.5911800 & 4493 $\pm$  20 & 3.113 $\pm$ 0.047 & 0.294 $\pm$ 0.022 &  -47 $\pm$  4 & 3854 $\pm$  52 & -0.110 $\pm$  0.018 &  7.89 $\pm$ 0.15 & 0.19941 $\pm$ 0.00917 & 0.7210 & Q1\\
    05089 & 0.4808195 & 2879 $\pm$  16 & 0.489 $\pm$ 0.050 & 0.630 $\pm$ 0.111 &  -67 $\pm$  7 & 6927 $\pm$  45 &  0.108 $\pm$  0.004 &  2.53 $\pm$ 0.88 & 0.00194 $\pm$ 0.00075 & 0.1032 & Q2\\
    05135 & 0.5191200 & 5795 $\pm$  85 & 3.888 $\pm$ 0.070 & 0.180 $\pm$ 0.013 &   51 $\pm$  3 & 3351 $\pm$  56 &  0.007 $\pm$  0.043 &  7.42 $\pm$ 0.06 & 0.23352 $\pm$ 0.00681 & 0.7826 & Q2\\
    05152 & 0.6201190 & 6645 $\pm$ 328 & 3.642 $\pm$ 0.194 & 0.352 $\pm$ 0.029 &  -51 $\pm$  3 & 4075 $\pm$  53 &  0.494 $\pm$  0.066 &  6.37 $\pm$ 0.12 & 0.14599 $\pm$ 0.00954 & 0.6163 & Q3\\
    05239 & 0.5230260 & 3288 $\pm$  41 & 0.303 $\pm$ 0.011 & 0.068 $\pm$ 0.049 & -127 $\pm$ 64 & 5335 $\pm$ 587 &  0.039 $\pm$  0.004 &  1.01 $\pm$ 0.04 & 0.00034 $\pm$ 0.00004 & 0.0556 & Q2\\
    05949 & 0.5007805 & 3285 $\pm$  15 & 2.109 $\pm$ 0.018 & 0.326 $\pm$ 0.015 &  -39 $\pm$  2 & 6073 $\pm$  21 & -0.104 $\pm$  0.011 &  7.39 $\pm$ 0.09 & 0.11596 $\pm$ 0.00286 & 0.5510 & Q2\\
    06498 & 0.5894900 & 2803 $\pm$   3 & 2.493 $\pm$ 0.010 & 0.136 $\pm$ 0.008 &  -77 $\pm$  3 & 6537 $\pm$  25 & -0.002 $\pm$  0.005 &  9.77 $\pm$ 0.04 & 0.26296 $\pm$ 0.00334 & 0.8332 & Q1\\
    06909 & 0.3731200 & 4204 $\pm$  47 & 1.443 $\pm$ 0.055 & 0.302 $\pm$ 0.067 &  136 $\pm$ 13 & 6526 $\pm$ 153 & -0.179 $\pm$  0.011 &  3.93 $\pm$ 0.20 & 0.02279 $\pm$ 0.00265 & 0.2678 & Q2\\
    06981 & 0.6076795 & 4195 $\pm$  35 & 0.900 $\pm$ 0.022 & 0.244 $\pm$ 0.032 & -162 $\pm$  7 & 5291 $\pm$  84 &  0.014 $\pm$  0.008 &  2.41 $\pm$ 0.08 & 0.00554 $\pm$ 0.00043 & 0.1528 & Q1\\
    06992 & 0.6075195 & 4897 $\pm$ 233 & 0.442 $\pm$ 0.037 & 0.252 $\pm$ 0.080 &  156 $\pm$ 18 & 4827 $\pm$ 247 &  0.121 $\pm$  0.028 &  1.02 $\pm$ 0.07 & 0.00048 $\pm$ 0.00010 & 0.0627 & Q3\\
    07051 & 0.8680320 & 2238 $\pm$  24 & 1.023 $\pm$ 0.038 & 0.284 $\pm$ 0.072 &  -32 $\pm$ 13 & 6856 $\pm$  83 &  0.469 $\pm$  0.024 &  5.21 $\pm$ 0.26 & 0.02860 $\pm$ 0.00317 & 0.2943 & Q2\\
    07079 & 0.5609450 & 6560 $\pm$ 277 & 1.839 $\pm$ 0.406 & 0.564 $\pm$ 0.039 &  170 $\pm$ 14 & 8114 $\pm$ 115 &  0.137 $\pm$  0.230 &  3.69 $\pm$ 0.76 & 0.02141 $\pm$ 0.01509 & 0.2609 & Q3\\
    07275 & 0.4594780 & 5535 $\pm$ 699 & 1.480 $\pm$ 0.209 & 0.299 $\pm$ 0.053 &  -12 $\pm$  5 & 6057 $\pm$  56 &  0.363 $\pm$  0.047 &  3.05 $\pm$ 0.10 & 0.01412 $\pm$ 0.00244 & 0.2203 & Q3\\
    07566 & 0.6768105 & 3517 $\pm$   8 & 1.745 $\pm$ 0.022 & 0.508 $\pm$ 0.013 &  177 $\pm$  1 & 5547 $\pm$  18 &  0.009 $\pm$  0.008 &  6.27 $\pm$ 0.12 & 0.05736 $\pm$ 0.00221 & 0.3979 & Q1\\
    07638 & 0.5553440 & 2285 $\pm$   9 & 1.251 $\pm$ 0.027 & 0.216 $\pm$ 0.043 & -159 $\pm$ 10 & 6535 $\pm$  63 &  0.067 $\pm$  0.015 &  6.11 $\pm$ 0.17 & 0.05011 $\pm$ 0.00325 & 0.3747 & Q1\\
    07640 & 0.5537705 & 1251 $\pm$   2 & 0.973 $\pm$ 0.011 & 0.076 $\pm$ 0.024 &  -54 $\pm$ 20 & 8293 $\pm$  71 &  0.060 $\pm$  0.004 &  8.49 $\pm$ 0.10 & 0.07854 $\pm$ 0.00275 & 0.4588 & Q1\\
    07659 & 0.5052320 & 4922 $\pm$ 131 & 0.350 $\pm$ 0.011 & 0.076 $\pm$ 0.047 &  -10 $\pm$ 40 & 7970 $\pm$ 572 & -0.078 $\pm$  0.008 &  0.78 $\pm$ 0.03 & 0.00024 $\pm$ 0.00002 & 0.0487 & Q3\\
    07943 & 0.6043500 & 3847 $\pm$  16 & 1.414 $\pm$ 0.018 & 0.174 $\pm$ 0.024 &  -25 $\pm$  8 & 6679 $\pm$  84 &  0.031 $\pm$  0.007 &  4.06 $\pm$ 0.06 & 0.02552 $\pm$ 0.00101 & 0.2806 & Q1\\
    07995 & 0.4784280 & 6330 $\pm$ 503 & 0.673 $\pm$ 0.055 & 0.398 $\pm$ 0.036 &   34 $\pm$  3 & 6373 $\pm$  42 &  0.079 $\pm$  0.008 &  1.26 $\pm$ 0.02 & 0.00101 $\pm$ 0.00009 & 0.0816 & Q3\\
    08185 & 0.4553630 & 4686 $\pm$  33 & 3.887 $\pm$ 0.024 & 0.270 $\pm$ 0.006 &  -50 $\pm$  1 & 5128 $\pm$  16 & -0.155 $\pm$  0.018 &  9.37 $\pm$ 0.05 & 0.35686 $\pm$ 0.00431 & 0.9840 & Q1\\
    08215 & 0.5031375 & 8819 $\pm$ 988 & 3.966 $\pm$ 0.671 & 0.279 $\pm$ 0.035 &   91 $\pm$  2 & 5408 $\pm$  38 &  0.429 $\pm$  0.155 &  5.07 $\pm$ 0.36 & 0.10815 $\pm$ 0.03036 & 0.5329 & Q3\\
    08442 & 0.5254550 & 3437 $\pm$  21 & 2.381 $\pm$ 0.028 & 0.290 $\pm$ 0.014 & -138 $\pm$  2 & 5827 $\pm$  25 & -0.211 $\pm$  0.027 &  7.87 $\pm$ 0.10 & 0.15243 $\pm$ 0.00499 & 0.6296 & Q1\\
    08697 & 0.5072340 & 2332 $\pm$  37 & 0.230 $\pm$ 0.037 & 0.581 $\pm$ 0.176 & -106 $\pm$ 10 & 7265 $\pm$  62 & -0.298 $\pm$  0.006 &  1.82 $\pm$ 2.45 & 0.00033 $\pm$ 0.00021 & 0.0544 & Q2\\
    08752 & 0.6124250 & 3930 $\pm$  19 & 2.273 $\pm$ 0.021 & 0.222 $\pm$ 0.017 &   26 $\pm$  4 & 6088 $\pm$  43 &  0.375 $\pm$  0.021 &  6.45 $\pm$ 0.07 & 0.10144 $\pm$ 0.00295 & 0.5169 & Q1\\
    08830 & 0.5891080 & 4800 $\pm$ 261 & 1.370 $\pm$ 0.089 & 0.512 $\pm$ 0.047 &   -8 $\pm$  4 & 5301 $\pm$  44 &  0.001 $\pm$  0.038 &  3.63 $\pm$ 0.29 & 0.01501 $\pm$ 0.00241 & 0.2258 & Q3\\
    09104 & 0.5150700 & 5438 $\pm$  47 & 1.501 $\pm$ 0.014 & 0.489 $\pm$ 0.011 & -103 $\pm$  1 & 6028 $\pm$  16 &  0.026 $\pm$  0.007 &  3.44 $\pm$ 0.03 & 0.01525 $\pm$ 0.00030 & 0.2272 & Q2\\
    09276 & 0.5138700 & 5179 $\pm$ 299 & 0.533 $\pm$ 0.072 & 0.731 $\pm$ 0.130 &  -98 $\pm$ 14 & 6700 $\pm$ 101 & -0.195 $\pm$  0.030 &  2.31 $\pm$ 3.13 & 0.00079 $\pm$ 0.00038 & 0.0747 & Q3\\
    09577 & 0.5598490 & 2150 $\pm$  10 & 1.274 $\pm$ 0.033 & 0.252 $\pm$ 0.058 & -114 $\pm$ 11 & 6237 $\pm$  62 & -0.926 $\pm$  0.020 &  6.67 $\pm$ 0.22 & 0.05975 $\pm$ 0.00470 & 0.4052 & Q1\\
    09635 & 0.5199447 & 3984 $\pm$ 210 & 0.792 $\pm$ 0.150 & 0.776 $\pm$ 0.072 & -175 $\pm$  5 & 7246 $\pm$  48 &  0.219 $\pm$  0.042 &  3.72 $\pm$ 1.72 & 0.00468 $\pm$ 0.00394 & 0.1433 & Q3\\
    09683 & 0.5992245 & 5383 $\pm$ 392 & 1.722 $\pm$ 0.133 & 0.278 $\pm$ 0.058 & -112 $\pm$ 13 & 6774 $\pm$ 156 &  0.077 $\pm$  0.057 &  3.63 $\pm$ 0.09 & 0.02355 $\pm$ 0.00243 & 0.2714 & Q3\\
    09698 & 0.4769498 & 5541 $\pm$ 522 & 0.566 $\pm$ 0.079 & 0.262 $\pm$ 0.082 &   66 $\pm$ 12 & 6487 $\pm$ 150 &  0.125 $\pm$  0.044 &  1.15 $\pm$ 0.08 & 0.00080 $\pm$ 0.00022 & 0.0750 & Q3\\
    09778 & 0.4900290 & 3095 $\pm$  55 & 0.247 $\pm$ 0.008 & 0.262 $\pm$ 0.053 &   84 $\pm$ 15 & 8767 $\pm$ 144 &  0.089 $\pm$  0.007 &  0.90 $\pm$ 0.03 & 0.00021 $\pm$ 0.00002 & 0.0467 & Q2\\
    09781 & 0.6973228 & 3454 $\pm$  93 & 0.199 $\pm$ 0.015 & 0.504 $\pm$ 0.084 & -170 $\pm$ 10 & 6411 $\pm$  87 &  0.055 $\pm$  0.006 &  0.74 $\pm$ 0.10 & 0.00009 $\pm$ 0.00002 & 0.0348 & Q2\\
    09789 & 0.6461960 & 3718 $\pm$   7 & 2.774 $\pm$ 0.021 & 0.222 $\pm$ 0.011 &   93 $\pm$  2 & 8470 $\pm$  25 &  0.308 $\pm$  0.010 &  8.32 $\pm$ 0.09 & 0.20605 $\pm$ 0.00505 & 0.7333 & Q1\\
    10047 & 0.4859560 & 4520 $\pm$ 323 & 0.405 $\pm$ 0.060 & 0.110 $\pm$ 0.074 &    3 $\pm$ 32 & 4567 $\pm$ 394 &  0.084 $\pm$  0.048 &  0.98 $\pm$ 0.08 & 0.00044 $\pm$ 0.00014 & 0.0607 & Q3\\
    10142 & 0.5082680 & 3734 $\pm$  18 & 2.120 $\pm$ 0.014 & 0.105 $\pm$ 0.007 &  129 $\pm$  4 & 5364 $\pm$  50 & -0.713 $\pm$  0.013 &  6.21 $\pm$ 0.03 & 0.09120 $\pm$ 0.00141 & 0.4917 & Q2\\
    10158 & 0.6122545 & 4413 $\pm$ 268 & 2.295 $\pm$ 0.141 & 0.378 $\pm$ 0.033 &   24 $\pm$  5 & 5634 $\pm$  51 & -0.071 $\pm$  0.109 &  6.12 $\pm$ 0.17 & 0.08294 $\pm$ 0.00700 & 0.4705 & Q3\\
    10210 & 0.5681330 & 3500 $\pm$  51 & 0.725 $\pm$ 0.023 & 0.487 $\pm$ 0.034 & -165 $\pm$  4 & 6838 $\pm$  32 & -0.011 $\pm$  0.017 &  2.59 $\pm$ 0.12 & 0.00416 $\pm$ 0.00036 & 0.1371 & Q2\\
    10356 & 0.4942865 & 4965 $\pm$ 366 & 2.614 $\pm$ 0.246 & 0.441 $\pm$ 0.027 &  114 $\pm$  4 & 6051 $\pm$  37 &  0.363 $\pm$  0.169 &  6.38 $\pm$ 0.22 & 0.09677 $\pm$ 0.01337 & 0.5056 & Q3\\
    10705 & 0.3599295 & 3427 $\pm$  28 & 0.864 $\pm$ 0.019 & 0.134 $\pm$ 0.034 & -133 $\pm$ 15 & 4241 $\pm$ 149 & -0.103 $\pm$  0.009 &  2.77 $\pm$ 0.06 & 0.00733 $\pm$ 0.00047 & 0.1702 & Q1\\
    10745 & 0.7294600 & 3768 $\pm$  29 & 1.073 $\pm$ 0.019 & 0.160 $\pm$ 0.030 &   62 $\pm$  9 & 8475 $\pm$ 105 &  0.069 $\pm$  0.013 &  3.14 $\pm$ 0.07 & 0.01160 $\pm$ 0.00063 & 0.2037 & Q1\\
    10906 & 0.4920660 & 3719 $\pm$  60 & 0.280 $\pm$ 0.009 & 0.308 $\pm$ 0.049 &  -63 $\pm$ 11 & 6409 $\pm$ 112 &  0.040 $\pm$  0.003 &  0.86 $\pm$ 0.03 & 0.00021 $\pm$ 0.00002 & 0.0469 & Q2\\
    11090 & 0.5940300 & 7618 $\pm$ 448 & 3.341 $\pm$ 0.307 & 0.201 $\pm$ 0.025 & -118 $\pm$  4 & 5192 $\pm$  76 &  0.244 $\pm$  0.100 &  4.86 $\pm$ 0.19 & 0.08609 $\pm$ 0.01387 & 0.4786 & Q3\\
    11098 & 0.5121868 & 5823 $\pm$ 506 & 0.470 $\pm$ 0.055 & 0.575 $\pm$ 0.071 &   25 $\pm$  8 & 5721 $\pm$ 106 & -0.012 $\pm$  0.030 &  1.08 $\pm$ 0.10 & 0.00042 $\pm$ 0.00011 & 0.0594 & Q3\\
    11105 & 0.5204120 & 4022 $\pm$  78 & 0.287 $\pm$ 0.012 & 0.507 $\pm$ 0.051 &  -47 $\pm$  9 & 7255 $\pm$  81 & -0.131 $\pm$  0.004 &  0.91 $\pm$ 0.06 & 0.00020 $\pm$ 0.00003 & 0.0457 & Q2\\
    11108 & 0.5983080 & 2877 $\pm$  28 & 0.685 $\pm$ 0.045 & 0.629 $\pm$ 0.119 & -107 $\pm$  6 & 8725 $\pm$  41 & -0.001 $\pm$  0.012 &  3.72 $\pm$ 2.40 & 0.00527 $\pm$ 0.00125 & 0.1499 & Q2\\
    11442 & 0.5145645 & 4739 $\pm$ 177 & 0.392 $\pm$ 0.010 & 0.508 $\pm$ 0.033 &  -17 $\pm$  4 & 7335 $\pm$  45 &  0.070 $\pm$  0.004 &  1.05 $\pm$ 0.03 & 0.00036 $\pm$ 0.00003 & 0.0564 & Q3\\
    11522 & 0.4216520 & 5114 $\pm$  24 & 0.817 $\pm$ 0.009 & 0.567 $\pm$ 0.015 &   11 $\pm$  1 & 8418 $\pm$  15 &  0.381 $\pm$  0.004 &  2.11 $\pm$ 0.05 & 0.00278 $\pm$ 0.00010 & 0.1180 & Q2\\
    11683 & 0.6346595 & 4999 $\pm$  51 & 2.375 $\pm$ 0.028 & 0.290 $\pm$ 0.017 &  -24 $\pm$  3 & 7045 $\pm$  44 & -0.007 $\pm$  0.013 &  5.40 $\pm$ 0.07 & 0.07154 $\pm$ 0.00241 & 0.4396 & Q1\\
    11730 & 0.7360930 & 5405 $\pm$ 288 & 3.355 $\pm$ 0.313 & 0.266 $\pm$ 0.083 &   45 $\pm$ 19 & 8433 $\pm$ 369 &  0.403 $\pm$  0.230 &  7.03 $\pm$ 0.36 & 0.17387 $\pm$ 0.03463 & 0.6724 & Q3\\
    11833 & 0.5625820 & 3238 $\pm$  11 & 1.865 $\pm$ 0.009 & 0.272 $\pm$ 0.010 &  -54 $\pm$  1 & 5911 $\pm$  15 &  0.039 $\pm$  0.011 &  6.51 $\pm$ 0.05 & 0.08251 $\pm$ 0.00121 & 0.4693 & Q1\\
    11966 & 0.5075527 & 6393 $\pm$ 337 & 0.394 $\pm$ 0.031 & 0.680 $\pm$ 0.047 &  -59 $\pm$  7 & 6976 $\pm$  65 &  0.036 $\pm$  0.014 &  0.92 $\pm$ 0.09 & 0.00020 $\pm$ 0.00003 & 0.0460 & Q3\\
    11989 & 0.5023055 & 4073 $\pm$  62 & 0.397 $\pm$ 0.017 & 0.312 $\pm$ 0.080 &   30 $\pm$ 13 & 7996 $\pm$ 149 &  0.234 $\pm$  0.007 &  1.12 $\pm$ 0.07 & 0.00050 $\pm$ 0.00007 & 0.0636 & Q2\\
    11990 & 0.4938388 & 1896 $\pm$  28 & 0.139 $\pm$ 0.007 & 0.199 $\pm$ 0.101 &   81 $\pm$ 39 & 6562 $\pm$ 203 &  0.107 $\pm$  0.005 &  0.82 $\pm$ 0.05 & 0.00010 $\pm$ 0.00002 & 0.0361 & Q2\\
    12333 & 0.6085633 & 3206 $\pm$  39 & 0.748 $\pm$ 0.019 & 0.061 $\pm$ 0.038 &  -87 $\pm$ 46 & 6378 $\pm$ 411 &  0.043 $\pm$  0.011 &  2.55 $\pm$ 0.07 & 0.00545 $\pm$ 0.00044 & 0.1519 & Q2\\
    12343 & 0.5415900 & 3275 $\pm$ 106 & 0.810 $\pm$ 0.074 & 0.445 $\pm$ 0.110 &   69 $\pm$ 17 & 4893 $\pm$ 185 &  0.247 $\pm$  0.042 &  3.05 $\pm$ 0.44 & 0.00672 $\pm$ 0.00165 & 0.1646 & Q3\\
    12466 & 0.4956910 & 4145 $\pm$ 102 & 0.359 $\pm$ 0.021 & 0.234 $\pm$ 0.102 &  100 $\pm$ 25 & 4562 $\pm$ 298 & -0.099 $\pm$  0.008 &  0.98 $\pm$ 0.10 & 0.00036 $\pm$ 0.00008 & 0.0567 & Q3\\
    12664 & 0.5169445 & 4476 $\pm$  96 & 0.834 $\pm$ 0.042 & 0.516 $\pm$ 0.049 & -179 $\pm$  6 & 6505 $\pm$  57 & -0.113 $\pm$  0.030 &  2.37 $\pm$ 0.16 & 0.00388 $\pm$ 0.00051 & 0.1335 & Q2\\
    12786 & 0.4816050 & 4884 $\pm$  80 & 0.556 $\pm$ 0.017 & 0.315 $\pm$ 0.045 &   17 $\pm$  7 & 8073 $\pm$  95 & -0.008 $\pm$  0.012 &  1.31 $\pm$ 0.04 & 0.00096 $\pm$ 0.00008 & 0.0800 & Q2\\
    12819 & 0.5102775 & 3698 $\pm$  89 & 0.267 $\pm$ 0.017 & 0.259 $\pm$ 0.071 & -146 $\pm$ 25 & 7724 $\pm$ 246 & -0.099 $\pm$  0.007 &  0.82 $\pm$ 0.06 & 0.00019 $\pm$ 0.00004 & 0.0449 & Q2\\
    13159 & 0.5451018 & 3381 $\pm$  54 & 0.692 $\pm$ 0.013 & 0.215 $\pm$ 0.039 &  149 $\pm$ 12 & 4710 $\pm$ 140 &  0.089 $\pm$  0.009 &  2.28 $\pm$ 0.05 & 0.00387 $\pm$ 0.00023 & 0.1334 & Q2\\
    13260 & 0.5689770 & 3231 $\pm$  11 & 0.775 $\pm$ 0.013 & 0.537 $\pm$ 0.026 &   92 $\pm$  3 & 7656 $\pm$  23 &  0.121 $\pm$  0.006 &  3.10 $\pm$ 0.09 & 0.00595 $\pm$ 0.00029 & 0.1571 & Q1\\
    13454 & 0.7109620 & 1076 $\pm$   6 & 0.764 $\pm$ 0.038 & 0.433 $\pm$ 0.100 &  132 $\pm$ 12 & 7969 $\pm$  44 &  0.053 $\pm$  0.022 &  8.67 $\pm$ 0.85 & 0.05173 $\pm$ 0.00841 & 0.3800 & Q1\\
    13477 & 0.5749995 & 4502 $\pm$  81 & 3.722 $\pm$ 0.090 & 0.139 $\pm$ 0.028 &  -30 $\pm$  8 & 5704 $\pm$ 100 & -0.010 $\pm$  0.075 &  9.08 $\pm$ 0.14 & 0.33949 $\pm$ 0.01569 & 0.9572 & Q2\\
    13534 & 0.4631050 & 3672 $\pm$  12 & 0.943 $\pm$ 0.011 & 0.271 $\pm$ 0.022 &   22 $\pm$  4 & 7455 $\pm$  41 &  0.003 $\pm$  0.004 &  2.90 $\pm$ 0.04 & 0.00830 $\pm$ 0.00030 & 0.1786 & Q1\\
    13896 & 0.6061850 & 3054 $\pm$  80 & 1.990 $\pm$ 0.088 & 0.295 $\pm$ 0.054 &  148 $\pm$  6 & 8687 $\pm$  66 & -0.051 $\pm$  0.024 &  7.43 $\pm$ 0.25 & 0.11289 $\pm$ 0.01087 & 0.5439 & Q2\\
    14101 & 0.5837305 & 4097 $\pm$  38 & 1.126 $\pm$ 0.041 & 0.423 $\pm$ 0.050 &  -21 $\pm$  7 & 7891 $\pm$  59 & -0.087 $\pm$  0.025 &  3.31 $\pm$ 0.21 & 0.01139 $\pm$ 0.00135 & 0.2022 & Q2\\
    14145 & 0.4727650 & 4066 $\pm$  12 & 2.303 $\pm$ 0.041 & 0.443 $\pm$ 0.021 &  -17 $\pm$  2 & 7148 $\pm$  30 &  0.252 $\pm$  0.006 &  6.88 $\pm$ 0.19 & 0.09868 $\pm$ 0.00522 & 0.5102 & Q1\\
    14526 & 0.6694710 & 2926 $\pm$  24 & 1.393 $\pm$ 0.058 & 0.389 $\pm$ 0.064 &   41 $\pm$  8 & 6185 $\pm$  66 &  0.152 $\pm$  0.028 &  5.65 $\pm$ 0.39 & 0.04236 $\pm$ 0.00549 & 0.3482 & Q1\\
    14784 & 0.5776860 & 2999 $\pm$ 227 & 0.849 $\pm$ 0.102 & 0.260 $\pm$ 0.128 & -169 $\pm$ 37 & 7038 $\pm$ 302 &  0.046 $\pm$  0.027 &  3.23 $\pm$ 0.39 & 0.00928 $\pm$ 0.00264 & 0.1866 & Q3\\
    14786 & 0.6100320 & 2332 $\pm$  56 & 1.290 $\pm$ 0.076 & 0.352 $\pm$ 0.070 & -106 $\pm$ 13 & 5237 $\pm$  91 & -0.396 $\pm$  0.102 &  6.46 $\pm$ 0.50 & 0.05299 $\pm$ 0.00832 & 0.3841 & Q3\\
    14815 & 0.5610100 & 2798 $\pm$  85 & 0.297 $\pm$ 0.035 & 0.477 $\pm$ 0.127 &  156 $\pm$ 15 & 6272 $\pm$ 117 & -0.173 $\pm$  0.014 &  1.36 $\pm$ 0.32 & 0.00047 $\pm$ 0.00020 & 0.0618 & Q3\\
    14830 & 0.6147847 & 4489 $\pm$ 186 & 0.989 $\pm$ 0.107 & 0.525 $\pm$ 0.141 &  157 $\pm$ 13 & 7007 $\pm$ 145 & -0.043 $\pm$  0.031 &  2.94 $\pm$ 0.68 & 0.00666 $\pm$ 0.00248 & 0.1640 & Q2\\
    14891 & 0.6368730 & 7525 $\pm$ 524 & 2.730 $\pm$ 0.430 & 0.169 $\pm$ 0.060 &  131 $\pm$ 17 & 8093 $\pm$ 309 &  0.259 $\pm$  0.299 &  3.99 $\pm$ 0.43 & 0.04939 $\pm$ 0.01817 & 0.3723 & Q3\\
    14905 & 0.5214227 & 3627 $\pm$  76 & 0.324 $\pm$ 0.033 & 0.447 $\pm$ 0.145 & -159 $\pm$ 13 & 7513 $\pm$ 131 & -0.006 $\pm$  0.009 &  1.13 $\pm$ 0.25 & 0.00036 $\pm$ 0.00013 & 0.0562 & Q2\\
    15388 & 0.5010780 & 2919 $\pm$  29 & 2.095 $\pm$ 0.085 & 0.407 $\pm$ 0.088 & -134 $\pm$  7 & 8269 $\pm$  63 &  0.361 $\pm$  0.028 &  8.63 $\pm$ 0.75 & 0.14483 $\pm$ 0.01943 & 0.6139 & Q3\\
    15394 & 0.4839540 & 3532 $\pm$  63 & 0.420 $\pm$ 0.018 & 0.093 $\pm$ 0.070 &  -65 $\pm$ 90 & 7498 $\pm$ 648 & -0.171 $\pm$  0.010 &  1.30 $\pm$ 0.07 & 0.00080 $\pm$ 0.00011 & 0.0748 & Q3\\
    15784 & 0.6415150 & 3657 $\pm$ 161 & 2.246 $\pm$ 0.144 & 0.286 $\pm$ 0.060 &  -93 $\pm$ 17 & 7813 $\pm$ 149 &  0.083 $\pm$  0.114 &  6.98 $\pm$ 0.25 & 0.11356 $\pm$ 0.01564 & 0.5455 & Q3\\
    15841 & 0.6605260 & 3411 $\pm$  83 & 2.104 $\pm$ 0.107 & 0.141 $\pm$ 0.062 &   -1 $\pm$ 35 & 4044 $\pm$ 359 &  0.528 $\pm$  0.045 &  6.79 $\pm$ 0.31 & 0.10735 $\pm$ 0.01427 & 0.5310 & Q2\\
    20376 & 0.7347950 & 1641 $\pm$  19 & 2.330 $\pm$ 0.313 & 0.628 $\pm$ 0.094 &  162 $\pm$  5 & 7657 $\pm$  25 &  3.687 $\pm$  0.179 & 20.75 $\pm$ 6.38 & 0.66507 $\pm$ 0.37041 & 1.4159 & Q3\\    20627 & 0.4444910 & 1186 $\pm$   9 & 0.261 $\pm$ 0.009 & 0.271 $\pm$ 0.074 &   95 $\pm$ 16 & 7507 $\pm$  54 &  0.114 $\pm$  0.014 &  2.49 $\pm$ 0.12 & 0.00169 $\pm$ 0.00018 & 0.0981 & Q1\\
    31312 & 0.4632650 & 2106 $\pm$  57 & 0.675 $\pm$ 0.025 & 0.265 $\pm$ 0.070 &  -29 $\pm$ 13 & 7528 $\pm$  76 & -0.074 $\pm$  0.084 &  3.63 $\pm$ 0.18 & 0.00930 $\pm$ 0.00092 & 0.1867 & Q2\\
    \enddata
    \tablecomments{The first two columns contain the OGLE ID (in the format OGLE-BLG-RRLYR-\textit{ID}) and the period used to construct the $O-C$ diagrams
    of the variables.
    The second set of six columns provides estimates for the orbital period, projected semimajor axis, eccentricity, argument of periastron,
    time of periastron passage (in the format HJD$-2{,}450{,}000$) and period-change rate, respectively. Note that these six parameters and their associated
    uncertainties are derived directly from their marginalized posterior distributions given by the employed MCMC method.
    The following set of three columns give estimates for the expected radial velocity semiamplitude, the mass function, and the minimum mass of the
    companion.
    The first two of these quantities are calculated for each of the posterior samples given by the MCMC analysis, and their
    means and standard deviations are listed here. The minimum masses of the companion objects are shown for the mean estimates of the mass functions
    of the variables only, and were calculated assuming $m_\mathrm{RR}\equiv 0.65\, M_\odot$ and $i \equiv 90$\degr.
    Finally, the last column indicates the quality class assigned to each of the variables, with Q1 being the most certain candidates for binary systems.}
\end{deluxetable*}

After the removal of the last few dubious cases, 87 stars remained on our final list.
For these stars, the final posterior samples were determined by running \textsc{emcee} for $31{,}000$ cycles
with 200 walkers, utilizing the priors listed in the Appendix. After an initial burn-in of 1000 iterations, the remaining chains samples
were thinned by a factor of 300, resulting in $20{,}000$ samples of the posterior probability distributions for each of the variables. 

Table~\ref{tab:binprop} presents the result of our analysis: for each of the 87 variables, it gives the OGLE identifier and the period
used to construct the $O-C$ diagram, along with estimates of various orbital parameters and derived quantities and their uncertainties.
The first two of these, $P_\mathrm{orb}$ and $a_1 \sin i$, are derived directly as the means and standard deviations of their marginalized
posterior distributions. The next two parameters, $e$ and $\omega$, are computed similarly, but only after transforming
the $\sqrt{e}\sin \omega$ and $\sqrt{e}\cos \omega$ values of the chain using Equations~\ref{eq:econv}~and~\ref{eq:omegaconv}. As $\omega$ is
a circular quantity, special care is taken to keep the posterior estimates contiguous for the calculation of its mean and standard
deviation, with the former quantity finally given in the range [$-180$\degr; $180$\degr].
The final two quantities directly derived from the LTTE + parabola fit, $T_0$ and the period-change rate
$\beta$ (after converting the $c_2$ values using Equation~\ref{eq:pchange}), are likewise given.

\begin{figure*}
    \includegraphics[width=\textwidth]{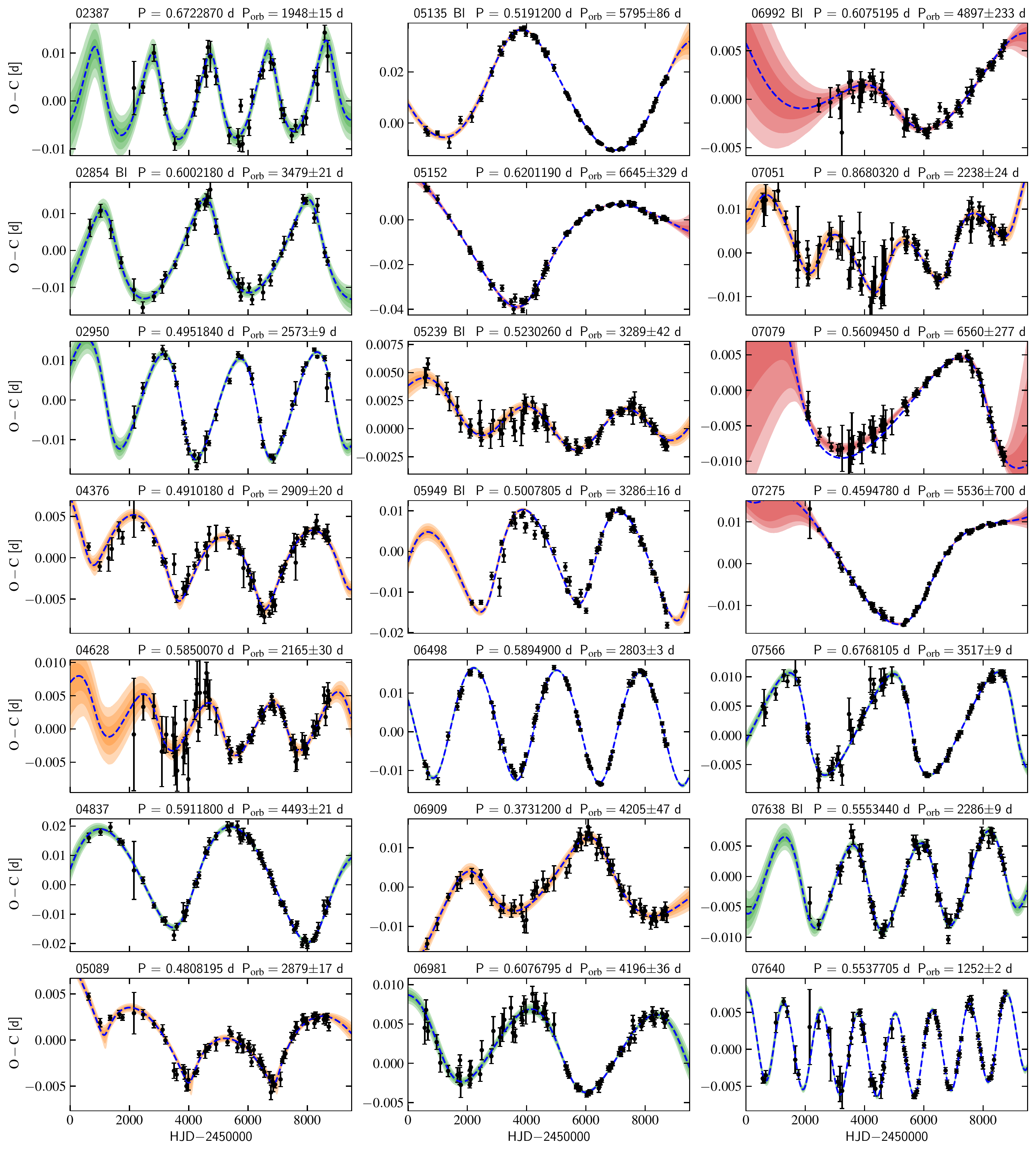}
    \caption{Final $O-C$ diagrams and their MCMC solutions. The shaded areas denote the ranges enclosing the fraction of
    posterior solutions corresponding to one, two, and three standard deviations (in order of decreasing transparency),
    while their color indicates which quality
    class has been assigned to the variable (green, yellow, and red for Q1, Q2, and Q3, respectively). The dashed blue line
    denotes the solution according to the mean parameters adopted from Table~\ref{tab:binprop}.
    The information given above each of the panels is the OGLE identifier of the RRL, the presence of the Blazhko effect
    in the light curve denoted by ``Bl'', the period used for constructing the $O-C$ diagram, and the orbital period inferred from its MCMC solution.
    \label{fig:ocs}}
\end{figure*}

\begin{figure*}\ContinuedFloat
    \includegraphics[width=\textwidth]{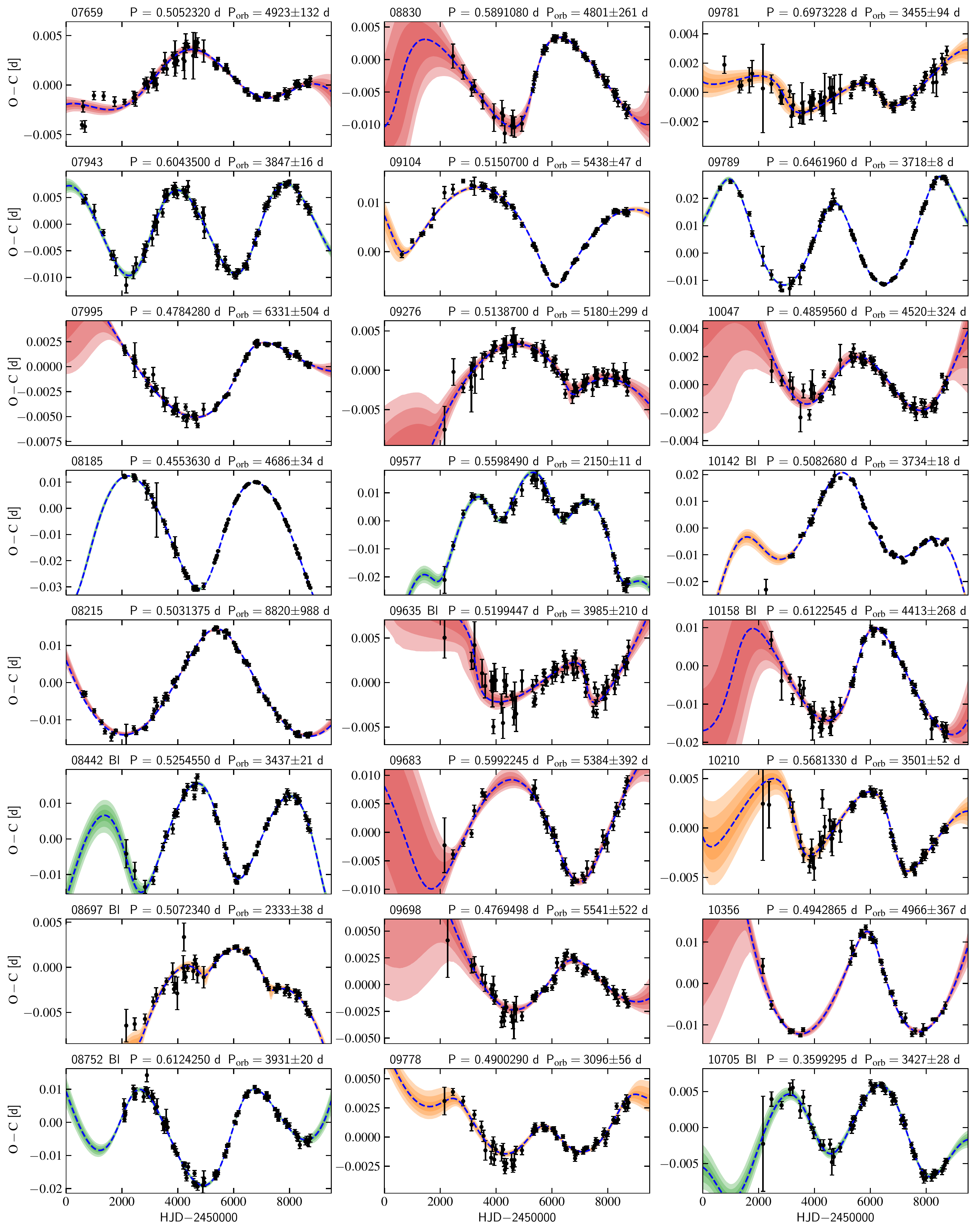}
    \caption{Continued from previous page.}
\end{figure*}

\begin{figure*}\ContinuedFloat
    \includegraphics[width=\textwidth]{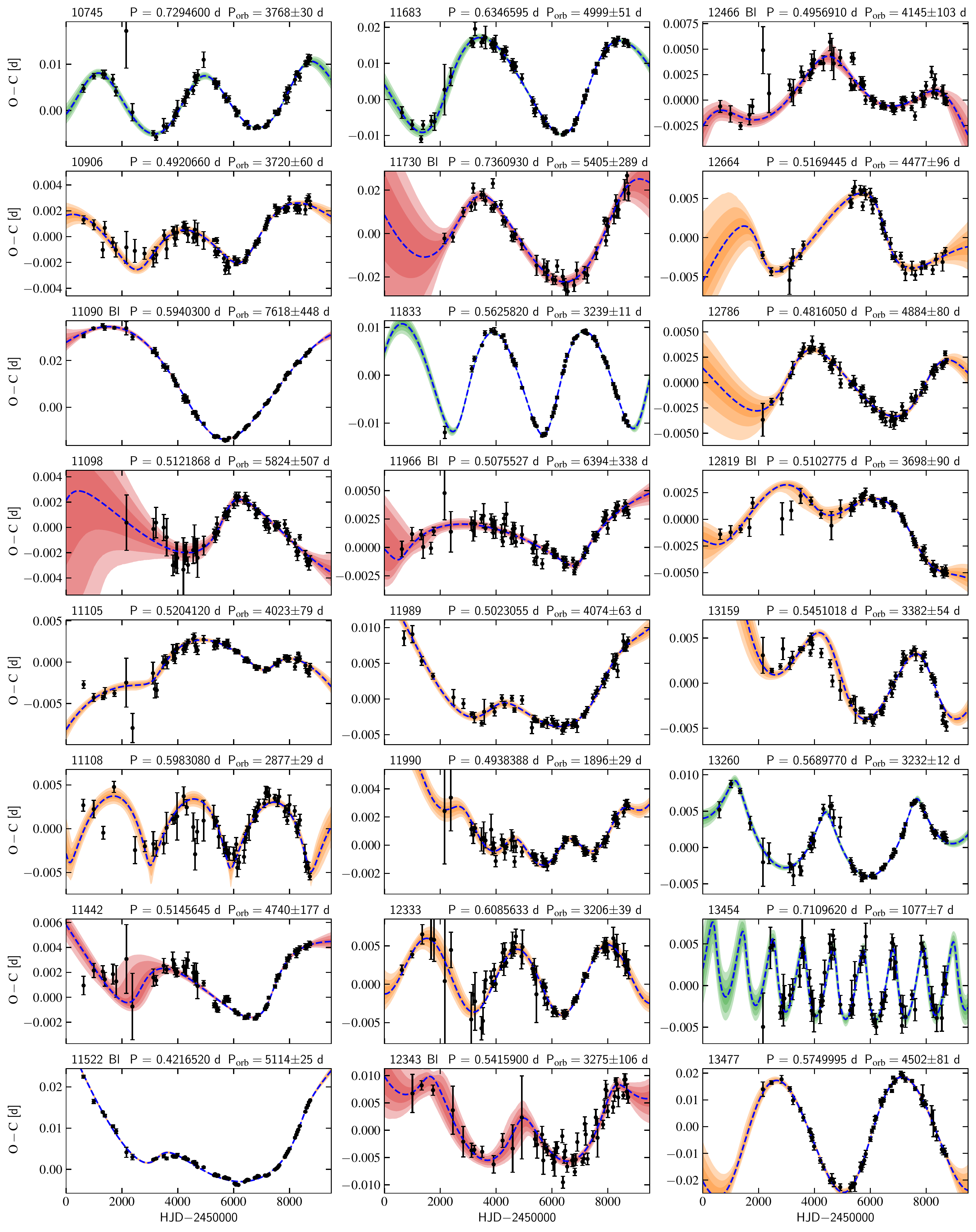}
    \caption{Continued from previous page.}
\end{figure*}

\begin{figure*}\ContinuedFloat
    \includegraphics[width=\textwidth]{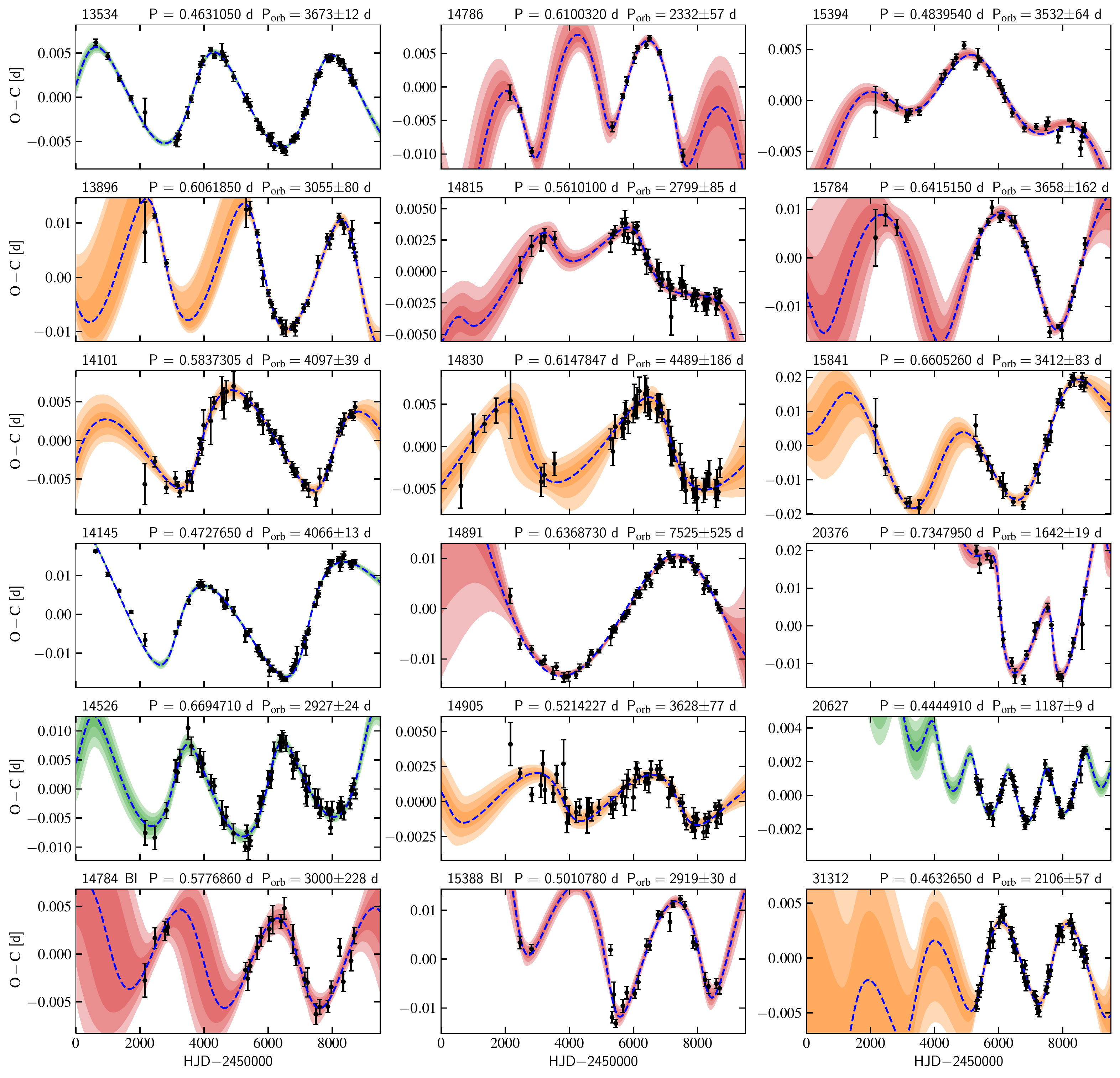}
    \caption{Continued from previous page.}
\end{figure*}

We also list some additional quantities derived from the orbital solutions.
The semiamplitude of the expected radial velocity variation of the RRL due to the binarity, $K_1$, can be expressed as

\begin{eqnarray}
    \label{eq:K}
    K_1 = \frac{2 \pi a_1 \sin i}{P_\mathrm{orb} \sqrt{1-e^2}},
\end{eqnarray}

\noindent which we have calculated for every posterior sample, with their averages and standard deviations being listed in Table~\ref{tab:binprop}.
As we have no additional information on the inclinations or mass ratios of the systems, the only physical quantity we can still calculate is the
mass function:

\begin{eqnarray}
    \label{eq:fm1}
    f(m) = \frac{a_1^3 \sin^3 i}{P_\mathrm{orb}\sqrt{1-e^2}},
\end{eqnarray}

\noindent which itself is connected to the masses of the components through

\begin{eqnarray}
    \label{eq:fm2}
    f(m) = \frac{m_\mathrm{S}^3 \sin^3 i}{(m_\mathrm{RR} + m_\mathrm{S})^2},
\end{eqnarray}

\noindent where $m_\mathrm{RR}$ and $m_\mathrm{S}$ are the masses of the RRL and the companion, respectively.
Similarly to the estimation of the expected radial velocity, we calculate the mean and standard deviation of the mass function after
calculating it for all samples of the posterior distributions, and these are listed in Table~\ref{tab:binprop}. Finally, we also list
the minimum masses of the RRL companions, calculated through Equation~\ref{eq:fm2} assuming $i \equiv 90\degr$ and
$m_{\rm RR}\equiv 0.65 M_\odot$, the latter being representative of typical values estimated with various methods
(see, e.g., \citealt{1993AJ....106..703S,1996ApJ...471L..33B} and \citealt{2018AJ....155..116C}, as well as references therein).

As the quality of the MCMC solutions varies by a great deal between different RRL variables, we have less confidence in some of our candidates
and therefore decided to indicate our degree of belief with an additional quality category, given in the last column of
Table~\ref{tab:binprop}. A quality category of Q1 marks the best candidates, and all of them have orbital periods with uncertainties less than $1\%$.
Candidates in category Q2 either have larger former relative errors than those of Q1, or there are discrepancies in their $O-C$ diagrams
that cause us to be more cautious with their classification.
The variables assigned a category of Q3 generally have problematic fits, in most cases due to the distribution or insufficient extent of the data,
which prevents the accurate determination of their binary parameters. For example, in the case of 20376, it has a pulsational
period-change rate four times larger than any other star in the sample, causing us to be cautious with it.
Nevertheless, we have decided to retain it, because the $O-C$ diagram extended with the last two
years of OGLE-IV observations follows exactly the trend extrapolated from the fit using data from the preceding years only.

Figure~\ref{fig:ocs} presents the $O-C$ diagrams of the 87 RRL binary candidates and their MCMC solutions, with dashed
lines denoting the binary solutions according to the average marginalized posterior estimates of the parameters given in Table~\ref{tab:binprop}.
Furthermore, the shaded areas give the ranges (credible intervals) containing $\sim 68.3\%$, $\sim 95.4\%$, and $\sim99.7\%$ of
$O-C$ solutions at any given point in time. The colors of these shaded areas denote the quality class assigned to a variable
(green, yellow, and red for Q1, Q2, and Q3 in Table~\ref{tab:binprop}, respectively).

\begin{deluxetable}{clc|clc}
    \tablecaption{RR~Lyrae Previously Classified as Binary Candidates \label{tab:previous}}
    \tablehead{
    \colhead{ID} & \colhead{References} & \colhead{Quality} &\colhead{ID} & \colhead{References} & \colhead{Quality}  \\ 
    }
    \startdata
    01573 & P19 & NB          & 09023 & P19 & NB\\
    02854 & P19 & Q1          & 09729 & P19 & NB\\
    03357 & P19 & NB          & 09778 & P19 & Q2\\
    04376 & H15; P19 & Q2     & 09789 & H15 & Q1\\
    04522 & H15 Table~1 & NB  & 10891 & H15 Table~1 & NB\\
    04628 & P19 & Q2          & 11522 & H15 & Q2\\
    05089 & P19 & Q2          & 11683 & H15 Table~1 & Q1\\
    05135 & H15 Table~1 & Q2  & 12333 & P19 & Q2\\
    05152 & H15 Table~1 & Q2  & 12611 & H15 Table~1 & NB\\
    05691 & H15; P19 & NB     & 12698 & H15 Table~1 & NB\\
    06498 & H15 & Q1          & 13454 & H18 & Q1\\
    06567 & P19 & NB          & 13477 & P19 & Q2\\
    06981 & H15; P19 & Q1     & 13534 & H15 & Q1\\
    07051 & P19 & Q2          & 14145 & H15 & Q1\\
    07566 & H15 & Q1          & 14408 & H15 & NB\\
    07640 & H15; P19 & Q1     & 14526 & P19 & Q1\\
    07943 & H15 & Q1          & 14852 & H15 Table~1 & NB\\
    08185 & P19 & Q1          & 14891 & P19 & Q3\\
    08215 & H18 & Q3          & 20627 & P19 & Q1\\
    \enddata
    \tablecomments{For each of the variables, the first two columns give the OGLE ID and the references to
    the work(s) where the star has been discussed previously. The references are:
    H15 -- \citet{2015MNRAS.449L.113H};
    H15~Table~1 -- \citet{2015MNRAS.449L.113H}, but no binary solution presented, and the star is only listed in Table~1 of that work;
    H18 -- \citet{2018pas6.conf..248H};
    P19 -- \citet{2019MNRAS.487L...1P}.
    The last column gives the same quality information as Table~\ref{tab:binprop}, except that stars that are no longer considered
    to be likely binary candidates are marked as ``NB.''}
\end{deluxetable}

\subsection{Previously announced candidates}\label{subsec:previous}

Previous studies have uncovered a total of 38 binary candidates toward the Galactic bulge through the
LTTE using $O-C$ diagrams.
In the original study of \citet{2015MNRAS.449L.113H} we have presented 20 potential candidates, and provided
binary parameters for 12 of them (eight more tentative candidates were listed in Table~1 of that study, with no
binary parameters determined for them).
Two more stars were presented as potential candidates in Fig.~2 of \citet{2018pas6.conf..248H}.
Finally, \citet{2019MNRAS.487L...1P} reported the discovery of 20 additional candidates; however, four stars
(04376, 05691, 06981, and 07640) had already been analyzed in \citet{2015MNRAS.449L.113H}, leaving them with
16 new prospective binaries.
All these variables were revised during our search, and the summarized findings are presented in Table~\ref{tab:previous}.

During our analysis, we have discarded 12 former candidates from our list of binary RRL candidates for different of reasons.
Most commonly, the $O-C$ points of stars deviate from the prediction based on the previously available shorter
data (01573, 06567, 09729), or the earlier and later parts of the $O-C$ diagram cannot be fit with
an LTTE solution consistently (10891, 12611, 12698, 14408, 14852). We attribute these cases mostly to random period changes
sometimes affecting RRL variables.
In other instances, the analysis of the light curve revealed that the apparent cyclic $O-C$ behavior is caused by the Blazhko effect
(04522, 05691).
Finally, we were unable to recover the periodic shape of the $O-C$ diagram presented by \citet{2019MNRAS.487L...1P} for two variables
(03357 and 09023), while employing our $O-C$ procedure described in Section~\ref{sec:search}.

\subsection{Blazhko effect and binarity}\label{subsec:blazhko}

Studies of the long-term (decades or longer) behavior of the Blazhko effect are few and far between, and generally concern
either individual, bright stars
(e.g., XZ~Dra: \citealt{2002A&A...396..539J}; RR~Gem: \citealt{2007A&A...469.1033S}; RV~UMa: \citealt{2007AN....328..841H};
RZ~Lyr: \citealt{2012MNRAS.423..993J}; RR~Lyr: \citealt{2014MNRAS.441.1435L})
or RRLs in globular clusters (Messier~5: \citealt{2011MNRAS.411.1763J}; Messier~3: \citealt{2012MNRAS.419.2173J}).
The general image that has emerged through these studies is that changes (including the appearance or disappearance) of
the Blazhko modulation are usually accompanied by changes in the pulsational period of the RRL (as traced with $O-C$ diagrams in most studies).
If these changes are (quasi)periodic, they can mimic the signal expected from binarity, as shown by \citet{2018MNRAS.474..824S} in the case of Z~CVn.
It should be mentioned, however, that there are Blazhko stars without sudden period changes on a time scale of $\sim 100$\,yr
(e.g. DM~Cyg, see Fig.~15 of \citealt{2009MNRAS.397..350J}).

\begin{deluxetable}{crr|rc}
    \tablecaption{Blazhko Periods Detected in the RR~Lyrae Binary Candidates  \label{tab:blazhko}}
    \tablehead{
    \colhead{ID} & \colhead{$P_\mathrm{Bl,1}$} & \colhead{$P_\mathrm{Bl, 2}$} & \colhead{$P_\mathrm{Bl, Ska.}$} & \colhead{Type}\\
    \colhead{}   & \colhead{(days)}            & \colhead{(days)}             & \colhead{(days)}                & \colhead{}     
    } 
    \startdata
    02854 &  18.50         &  67.01 &     -- & -- \\ 
    05135 & \textit{12.63} &     -- &     -- & -- \\ 
    05239 &  10.68         &     -- &     -- & -- \\ 
    05949 &  21.60         &     -- &     -- & -- \\ 
    06992 & 354            &     -- &     -- & -- \\ 
    07638 & 234.54         &     -- & 235.29 &  c \\ 
    08442 &  19.31         & 167.45 &  19.33 &  a \\ 
    08697 & \textit{31.72} &     -- &     -- & -- \\ 
    08752 &  27.30         &  51.50 &  27.29 &  a \\ 
    09635 &  22.94         &     -- &  22.94 &  a \\ 
    10142 &  23.52         &     -- & 238.10 & -- \\ 
    10158 &  23.25         &     -- &  23.23 &  c \\
    10705 &   7.01         &     -- &   7.01 &  a \\
    11090 &  290.6         &     -- &     -- & -- \\
    11522 &  16.14         &     -- &     -- & -- \\
    11730 &  54.08         &     -- &  54.05 &  a \\
    11966 &  26.33         &     -- &     -- & -- \\
    12343 & 102.23         & 175.30 & 102.04 & -- \\
    12466 &  23.16         &  17.28 &  23.15 &  a \\
    12819 &  23.13         &     -- &     -- & -- \\
    14784 &  20.62         &     -- &     -- & -- \\
    15388 & 145.80         &     -- &     -- & -- \\
    \enddata
    \tablecomments{For each of the variables, the meaning of the columns in order is: the OGLE ID;
    the first detected Blazhko period; the second detected Blazhko period; the Blazhko period reported
    by \citet{2020MNRAS.494.1237S}; and the latter authors' reported morphological type
    (both only if present in their Table~1).
    Blazhko periods with high uncertainty are marked in \textit{italics}.}
\end{deluxetable}

Besides affecting the pulsational period of RRL stars, the phase modulation of the light curve caused by the Blazhko effect can also be
misinterpreted as a signal of binarity.
As the modulation periods reach well into the range of expected orbital periods for RRL
binaries,\footnote{V144 in Messier~3 has a Blazhko period longer than 25\,yr \citep{2016CoKon.105..167J}.}
this is especially problematic in the case of studies making use of only timings of maxima.
This is the exact reason why, as mentioned in Section~\ref{subsec:initial}, we have inspected the period change plus LTTE-corrected
light curves of each of the variables for signs of the Blazhko effect, and discarded those candidates whose detected Blazhko periods
were similar to the suspected orbital period.
This search (based mostly on the inspection of Fourier spectra of residual light curves for side peaks of the main harmonics of the
pulsational frequency, as well as folded light curves and animations of the suspected modulation periods)
has also revealed that 22 of the 87 candidates indeed exhibit the Blazhko effect (besides their modulation from LTTE).
The modulation periods found are listed in Table~\ref{tab:blazhko}, alongside their values and modulation classes (when present)
previously published by \citet{2020MNRAS.494.1237S}. Generally, we find the same dominant modulation period, except for 10142,
which admittedly has a somewhat complicated residual power spectrum, with multiple plausible periods.
For five stars, we have also found a significant secondary modulation signal.

The frequency of RRLs showing the Blazhko effect is estimated to be $\sim50\%$ \citep{2009MNRAS.400.1006J, 2010ApJ...713L.198K,2019ApJS..244...32P},
although recently there have been suggestions that all RRLs are in fact modulated, but a significant fraction are modulated with very small amplitudes \citep{2018A&A...614L...4K}.
The presence of the Blazhko effect can hinder the detection of the LTTE through its well-documented connection to
sudden period changes, or by having a long-term phase modulation component, preventing the decoupling of the two effects. 
There are 65 stars where we do not detect its presence. Hence, if we adopt $50\%$ as its incidence rate in RRL variables, we should have
detected $\sim130$ stars with LTTE. This also means that for variables afflicted by the presence of the Blazhko effect, the
LTTE signal can still be recovered in $\sim 33\%$ of the cases. As the Blazhko effect prevents the detection of the LTTE signal in the
remaining $\sim66\%$ of affected variables, we can surmise that the binary nature of Blazhko variables can only be recovered in one third of cases.

\section{Statistical properties of the candidate RR~Lyrae binary population}

\begin{figure}
    \includegraphics[width=0.45\textwidth]{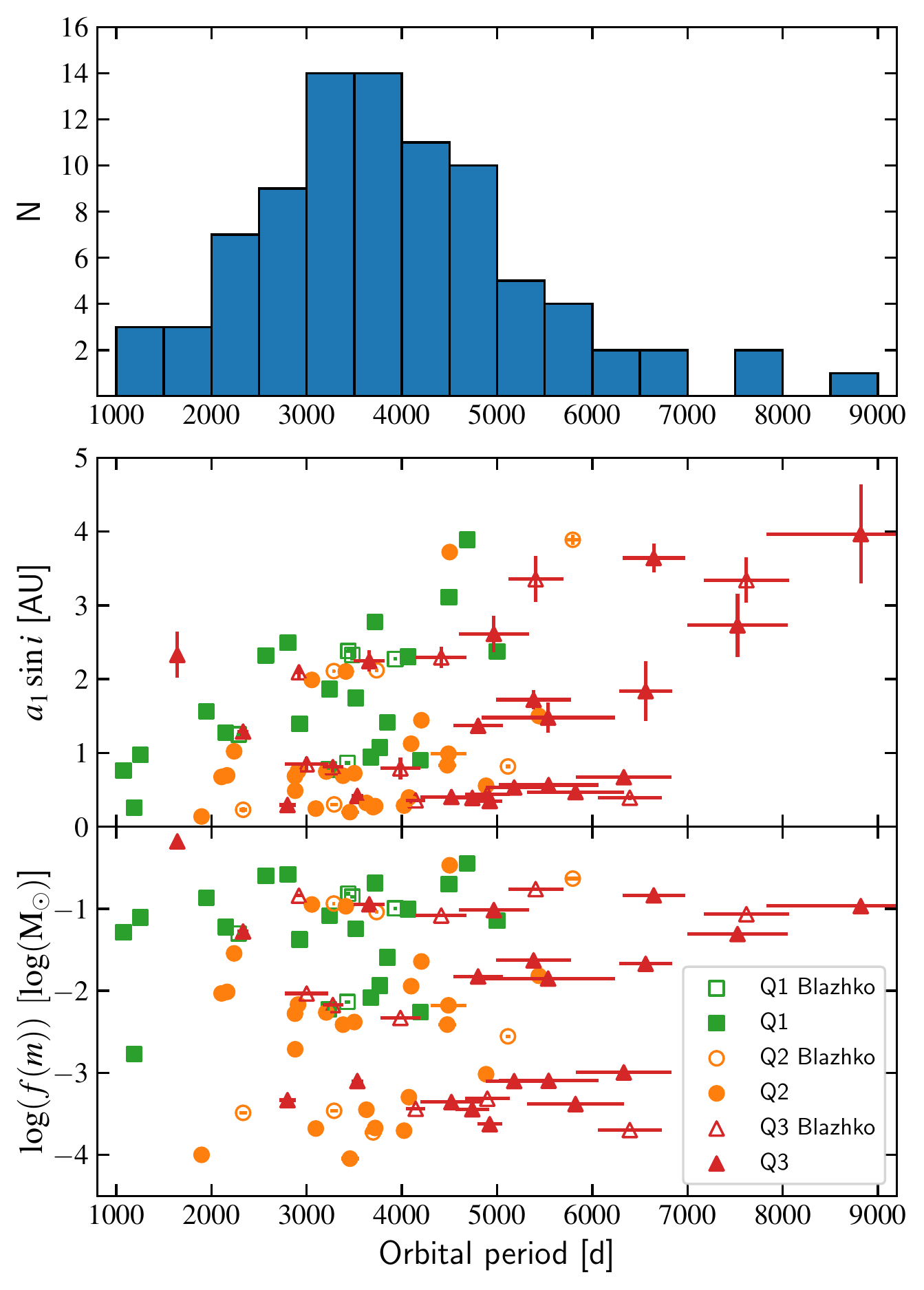}
    \caption{Top: orbital period distribution of the bulge RRL binary candidate sample.
    Middle: the distribution of the projected semimajor axes as a function of the orbital periods.
    Green squares, yellow circles, and red triangles denote variables with assigned quality
    of first, second, and third class, respectively, as listed in Table~\ref{tab:binprop}.
    Empty symbols indicate variables with presence of the Blazhko effect (Table~\ref{tab:blazhko}).
    Bottom: same as the middle panel, but for the distribution of the (logarithmic) mass function from
    Table~\ref{tab:binprop}.
    \label{fig:period}}
\end{figure}

For the first time, the sizable list of candidates presented in this work allows us to probe the population of
RRLs in binaries, as well as to draw conclusions on their companion objects.
The top panel of Figure~\ref{fig:period} illustrates the distribution of orbital periods for the complete sample.
The minimum period of $\sim1000$ days is a hard limit on these objects, because not a single one has been detected by our
search below this limit. It should be noted that with the method described in Section~\ref{sec:search}, we should have detected
companions with orbital periods below 1000\,days, if they existed.
The frequency of binaries steadily increases from this limit up to periods of $\sim3000 - 4000$\,days,
then decreases toward even longer periods.
As the OGLE project evolved, the list of observed fields has been changed, sometimes resulting in shorter
RRL light curves than the total possible baseline, mostly in OGLE-III.\footnote{This can be clearly seen in the $O-C$ diagrams
in Figure~\ref{fig:ocs} for, e.g, 13896, 14784, 14786, and 14815, among others.} Gaps in the light curves prevent the discovery
of longer-period binary RRLs in many cases, where it can be hard to decide whether the structure of the $O-C$ diagram
is caused by the LTTE or by changes in the pulsation period. This is especially true for variables with low LTTE amplitudes,
which, as described in Section~\ref{subsec:bin}, were discarded when it was not clear whether random period changes or the LTTE
itself was causing the $O-C$ shapes.
As can be seen in the middle panel of Figure~\ref{fig:period}, illustrating the distribution of semimajor axes,
long-period, low-amplitude binaries are absent from the sample as a consequence.
Furthermore, it can also be seen on this diagram  that the maximum values of the projected semimajor axes seemingly increase with increasing periods.
As illustrated by the distribution of mass-function values in the bottom panel of Figure~\ref{fig:period}, this roughly
equates to an upper limit for the mass function, which in turn hints at a limit for the companion masses of RRL binaries.

\begin{figure}
    \includegraphics[width=0.45\textwidth]{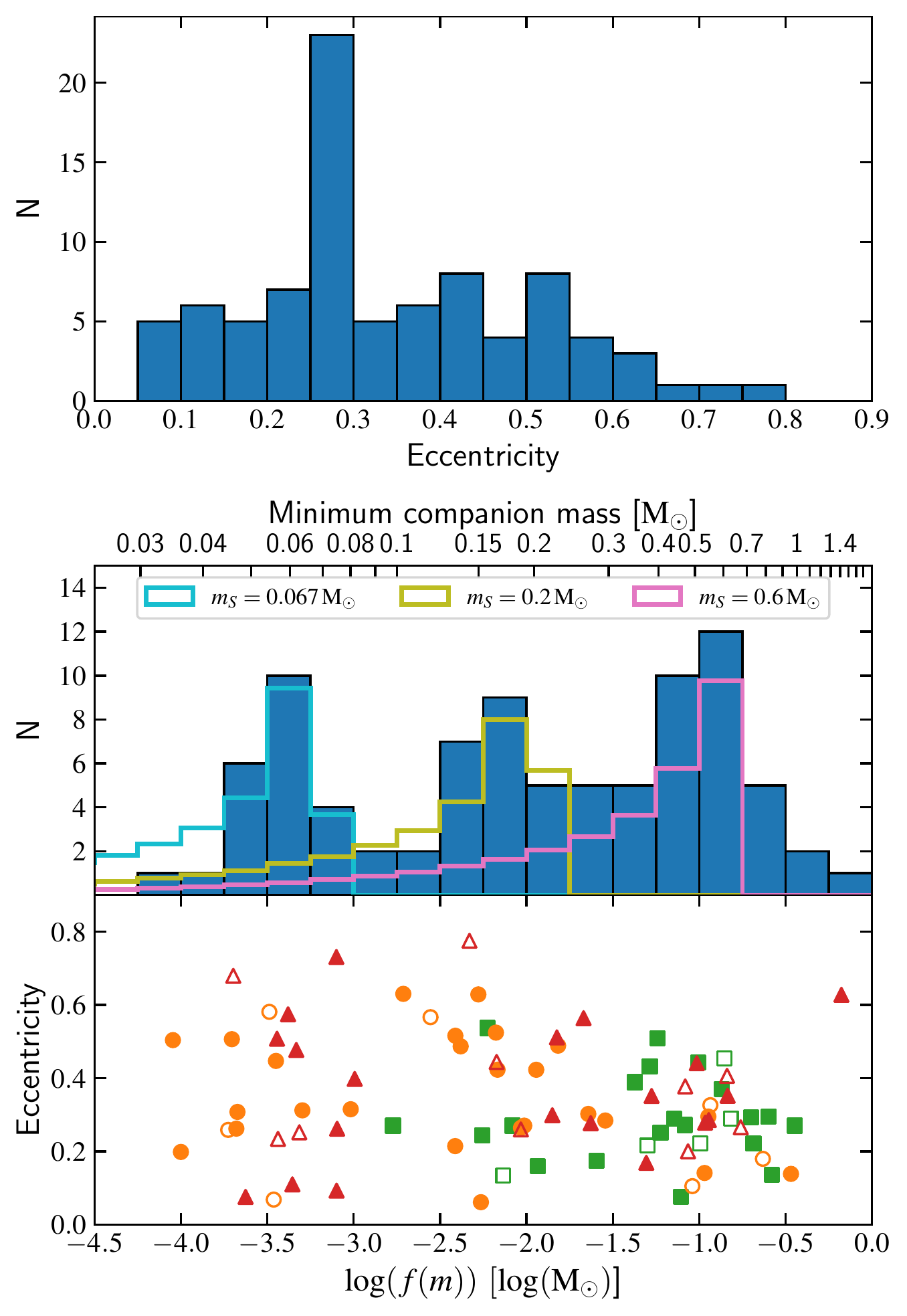}
    \caption{Top: eccentricity distribution of the bulge RRL binary candidate sample.
    Middle: distribution of the (logarithmic) mass function. In addition, the
    purple, yellow, and cyan lines show the expected mass-function distribution for companion masses of
    0.6, 0.2, and $0.067M_\odot$, respectively, selected to reproduce the three modes of the observed sample.
    Note that the top axis shows the minimum companion mass calculated for an assumed RRL mass of $0.65\,M_\odot$. 
    Bottom: the dependence of eccentricities on the (logarithmic) mass function in the bulge RRL binary sample.
    The meaning of the symbols is the same as in Figure~\ref{fig:period}.
    \label{fig:fm}}
\end{figure}

The top panel of Figure~\ref{fig:fm} illustrates the distribution of eccentricities for the RRL binary candidate sample.
Although the uncertainties on individual values are somewhat high for some of the candidates (as listed in Table~\ref{tab:binprop}),
random fluctuations cannot explain the most prominent visible feature, the peak in the bin $0.25 < e < 0.3$.
This bin contains $\sim4$\,times as many candidates as the average for other bins between 0.05 and 0.6. Supposing a Poisson
distribution with a mean measured in these bins, the probability of a single bin (i.e., that of $0.25 < e < 0.3$) having more than
20 systems is less than $10^{-6}$.
Furthermore, systems with eccentricities higher than 0.6 appear to be exceedingly rare (only six candidates, or 7\% of the total sample).

The distribution of the derived mass-function values for the sample of RRL binary candidates is shown in the middle panel of
Figure~\ref{fig:fm} (on a logarithmic scale). The distribution is strikingly trimodal, strongly suggesting the presence of three
dominant underlying populations of companions to RRL variables in binary systems.
While we cannot derive the masses of individual companions from the information provided by LTTE alone, we can calculate the expected
distribution of their $f(m)$ values through Equation~\ref{eq:fm2}, by assuming the same mass for all RRL components (adopted here as
$0.65\,M_\odot$, see Section~\ref{sec:parameters}), and an isotropic inclination distribution. We have found that under these conditions,
dominant companion masses of
$\sim0.6$, $\sim0.2$, and $\sim0.067\,M_\odot$ are able to reproduce the location, and to a lesser extent the shape, of
these three peaks. It should be noted that if the adopted mean RRL mass is decreased to $0.55\,M_\odot$, or increased
to $0.75\,M_\odot$, the dominant companion masses likewise decrease or increase by about $\sim10\%$ for all three groups.

The bottom panel of Figure~\ref{fig:fm} illustrates the relation between the derived eccentricities and mass functions for the
RRL binary candidates. While the low number of candidates complicates interpretation, it can still be surmised that all three
mass-function groups contribute to the excess of variables with $0.25<e<0.3$. The group with the
highest $f(m)$ values seems to have  a roughly uniform underlying distribution between eccentricities of $\sim0.1$ and $\sim0.5$.
In contrast, the groups with lower $f(m)$ values show a paucity of eccentricity values around $\sim0.4$ and a less
significant excess around $\sim0.5$.

\section{Discussion} \label{sec:discussion}

As already demonstrated in this study, the analysis of the $O-C$ diagram, when coupled with the modified \citet{1919AN....210...17H} method 
presented in Section~\ref{sec:search}, is a powerful tool for the discovery of RRL binary candidates.
Inspecting the presented $O-C$ diagrams, we can notice some general trends. For example, in the case of some variables
(04376, 06498, and 09577) with more than two completed orbits, at some time intervals the $O-C$ points are systematically above
the $O-C$ solutions while at others they are below. This phenomenon is most probably caused by the nonlinear change in
the pulsational periods of the variables themselves. With continued observations, it will be possible to model this
by extending the polynomial part of Equation~\ref{eq:parabola} to higher orders, as was done by \citet{2017A&A...603A..70G}
in the case of Type~II and anomalous Cepheids.

As we have adopted the Bayesian formalism with MCMC analysis, we can evaluate possible correlations between the binarity parameters
using the distribution of posterior probabilities. Most candidates with long binary periods compared to the length of the available data
(i.e., 07079, 07995, 08830, 09683, etc.) show diverging solutions (the shaded areas) in their $O-C$ diagrams.
This is the result of a degeneracy between the pulsational period-change rates, the orbital periods, and the projected semimajor axis, because a variety of
combinations of these three parameters can reproduce the same shape for short segments of the $O-C$ curves (Equation~\ref{eq:parabola}).
Although this prevents the derivation of accurate binary parameters, reflected in the classification of these stars as quality category Q3,
as well as in the large relative errors given for these parameters in Table~\ref{tab:binprop},
the characteristic shape of the LTTE on the $O-C$ diagrams still allows their identification as binary candidates.

Another advantage of the adopted Bayesian formalism (albeit not yet used in this work) is that it can naturally include
external information through the appropriate selection of priors. For example, if any of our candidates were to show even a single eclipse,
the time of the eclipse could be converted to a prior through the times of inferior ($\omega + \nu=90^\circ$) or superior
($\omega + \nu=270^\circ$) conjunctions, providing a strong constraint on the binary parameters. The best known
example for such a system is the binary Cepheid OGLE-SMC-CEP-3235 (see Fig.~2 in \citealt{2015AcA....65..341U}),
showing both a clear LTTE in its $O-C$ diagram and a single eclipse.

\subsection{The nature of RRL companions} \label{subsec:comp}

As RRL stars are Population~II objects, the properties of their companions can provide important information about star formation
in conditions no longer observable in the Milky Way.
The distribution of mass-function values presented in Figure~\ref{fig:fm} provides strong clues toward this goal.
There are three RRL binary candidates with estimated minimum companion masses higher than $0.95M_\odot$, that is,
similar to or higher than the expected masses of main-sequence progenitors of RRL variables. These systems are the current
best candidates for RRL binaries containing high-mass degenerate companions.
\citet{2013MNRAS.435..698S} modeled wind-driven accretion of material enhanced in carbon and neutron-capture elements
onto stars that eventually form RRL variables from their higher-mass, more evolved AGB companions.
This scenario could be verified if some of our RRL binary candidates were to show enhancement of these elements in their spectra.
Whether there would be a chemical signature associated with neutron star or black hole companions should also be investigated.
Alternatively, a more mundane scenario with hierarchical triple systems, containing two low-mass main-sequence companions in a tight binary,
accompanied by the RRL star on a wider orbit, can also explain these high minimum companion masses.

As shown in Figure~\ref{fig:fm}, the three modes of the mass-function distribution can be modeled with typical companion masses belonging
to three distinct companion populations.
In the globular cluster Messier~4 (NGC~6121), which itself contains a sizable population of RRL variables, the youngest single white dwarfs have masses of
$\sim0.53M_\odot$ \citep{2009ApJ...705..408K}. This mass constitutes a lower limit on the masses of degenerate companions to RRL variables,
under the assumption of no interaction between the components in the system. Older white dwarfs, the remnants of initially
(at the formation of the system) higher-mass companions, are expected to follow the initial--final mass relation (IFMR; \citealt{2018ApJ...866...21C}).
The IFMR itself is quite flat, i.e., the final masses of white dwarfs around $\sim0.6M_\odot$ change little for a relatively wide range
of initial masses (e.g., $\sim0.57\,M_\odot$ and $\sim0.65\,M_\odot$ for $1\,M_\odot$ and $2\,M_\odot$, respectively,
according to Equation~4 in \citealt{2018ApJ...866...21C}). The white dwarf mass distribution observed in the field supports this argument
\citep{2007MNRAS.375.1315K}.
Hence, we surmise that the peak corresponding to the highest-mass companions in Figure~\ref{fig:fm}
probably contains a significant population of white dwarf remnants of initially higher-mass (1--2$\,M_\odot$) companions,
with the rest having companions still on the main sequence. However, the relative fraction of these two companion populations
is not yet clear and will require further study.

White dwarfs with extremely low masses (ELMs; $\sim0.2\,M_\odot$) have been observed in the field \citep{2004ApJ...606L.147L}
and in binary systems \citep{2019ApJ...881L...3M}, and are generally thought to form through binary interaction. Therefore, it is tempting
to attribute the middle-mass group ($\sim0.2\,M_\odot$) of RRL companions seen in Figure~\ref{fig:fm} to these objects.
However, considering the long binary periods of these systems, the RRL progenitor probably did not interact with its companion. Instead, 
to facilitate the binary interaction scenario, a third object is required in the system in an initially close orbit with the supposed ELM progenitor.
However, this third body would contribute its mass to that of the ELM to produce an LTTE bigger than observed in the light curve of the RRL variables,
unless it was somehow ejected from the system. Alternatively, companions in this mass group might be ordinary M dwarfs, signaling a
peculiar distribution distribution of mass ratio ($q$) in Population~II binaries, with a strong peak at $q\sim0.2$.
This would mean that the mass ratio distribution for low-mass Population~II binaries is very different from the mostly smooth and
flat distribution observed for solar-type stars in the solar neighborhood for similar orbital periods (see fig.~30 of \citealt{2017ApJS..230...15M}).

The existence of the low-mass group (located at ${\sim0.067}\,M_\odot \approx 70\,M_\mathrm{Jup}$) is arguably the most puzzling feature of the
mass-function distribution. These candidate companions are situated firmly in the brown dwarf (BD) mass regime
(${\lesssim0.075} M_\odot$; \citealt{1997ApJ...491..856B}).  Population~II BDs are very faint objects, and as such, very few have been detected
in the Galactic field (see, e.g., \citealt{2017MNRAS.468..261Z}). Furthermore, very few extended radial velocity surveys of
low-metallicity stars have been carried out, with the notable exception of the APOGEE observations \citep{2016AJ....151...85T,2020ApJ...895....2P}.
Nevertheless, most of the BD candidates detected in APOGEE have much shorter orbital periods than the ones presented here.
Therefore, the detection of RRL binary candidates containing  these objects may provide us with valuable information on the binarity
of Population~II BDs, which is not easily attainable through other methods.

Regardless of the nature of a companion, a binary system containing an RRL variable has necessarily undergone at least some evolution
in its parameters as a consequence of mass loss on the red giant branch. Considerable effort toward binary evolution modeling and
population synthesis studies are needed to recover the distributions of the original stellar and binary parameters of these objects.
Nevertheless, they offer us a rather unique look into properties of the slightly subsolar mass stars of Population~II.

On the observational side, continued photometric monitoring will result in extended $O-C$ diagrams and more accurate binary parameters.
However, radial velocity observations are still necessary to verify at least a subset of the candidates of each $f(m)$ group, before they
can be used to draw firm conclusions about the binary parameters of Population~II. This will be especially
challenging for the candidates with the lowest-mass companions, because the expected radial velocity semiamplitudes listed in Table~\ref{tab:binprop}
are $\sim1\,\mathrm{km\,s}^{-1}$. These are overlaid on the large radial velocity variations caused by the pulsation of RRL stars
(up to $\sim60\,\mathrm{km\,s}^{-1}$; see e.g., Fig.~3 of \citealt{2017MNRAS.468.1317J}), hence the verification of these binary systems poses
considerable challenge.

\acknowledgments
The research leading to these results has received funding from the European Research Council (ERC) under the European Union's Horizon 2020
research and innovation program (grant agreement No.~695099). We also acknowledge support from the National Science Center, Poland grants
MAESTRO UMO-2017/26/A/ST9/00446, MAESTRO 2016/22/A/ST9/00009 and DIR/WK/2018/09 grants of the Polish Ministry of Science and Higher Education.
The OGLE project has received funding from the National Science Centre, Poland, grant MAESTRO 2014/14/A/ST9/00121 to AU.
J.J. acknowledges support from the OTKA grants NN-129075 and K-129249.
Support for M.C. is provided by the Ministry for the Economy, Development, and Tourism's Millennium Science Initiative through grant
IC\,12009, awarded to the Millennium Institute of Astrophysics (MAS) by Proyecto Basal AFB-170002; and by FONDECYT grant \#1171273.

\vspace{5mm}
\facilities{OGLE.}

\software{
          emcee \citep{2013PASP..125..306F}, 
          ChainConsumer \citep{Hinton2016},
          Jupyter \citep{jupyter},
          Numba \citep{Numba},
          NumPy \citep{harris2020array},
          SciPy \citep{2020SciPy-NMeth},
          scikit-learn \citep{scikit-learn}
          }




\appendix

\restartappendixnumbering

\section{Priors utilized during the MCMC analysis} \label{app:priors}

\begin{deluxetable*}{c|cc|cc|cc||c|cc|cc|cc}
    \tablecaption{Fourier orders and priors used to derive the final RRL binary parameters\label{tab:priors}}
    \tablewidth{700pt}
    \tabletypesize{\scriptsize}
    \tablehead{
    \colhead{ID} & 
    \colhead{$\mathcal{F}_1$} & \colhead{$\mathcal{F}_2$} &  \colhead{$P_\mathrm{orb,min.}$} & \colhead{$P_\mathrm{orb,max.}$} &
    \colhead{T$_\mathrm{0,min.}$} & \colhead{T$_\mathrm{0,max.}$} &
    \colhead{ID} & 
    \colhead{$\mathcal{F}_1$} & \colhead{$\mathcal{F}_2$} &  \colhead{$P_\mathrm{orb,min.}$} & \colhead{$P_\mathrm{orb,max.}$} &
    \colhead{T$_\mathrm{0,min.}$} & \colhead{T$_\mathrm{0,max.}$}
    } 
    \startdata
    02387 &  8 & 10 & 1766.6 & 2159.2 & 8295.3 & 9276.7 & 10705 & 10 & 15 & 3085.0 & 3564.9 & 3800.0 & 5000.0\\
    02854 & 10 & 15 & 3133.3 & 3829.6 & 7345.0 & 9085.7 & 10745 & 15 & 20 & 3272.3 & 3999.4 & 7293.3 & 9111.2\\
    02950 & 10 & 15 & 2295.5 & 2805.7 & 5764.0 & 7039.3 & 10906 & 10 & 15 & 3333.1 & 4073.7 & 5830.4 & 7682.1\\
    04376 & 10 & 15 & 2610.6 & 3190.8 & 5798.9 & 7249.2 & 11090 & 10 & 15 & 5599.2 & 8398.8 & 3414.1 & 6913.6\\
    04628 & 10 & 15 & 1887.0 & 2306.3 & 6723.1 & 7553.0 & 11098 & 10 & 15 & 4188.4 & 6515.3 & 4368.9 & 7083.6\\
    04837 & 10 & 15 & 4058.6 & 4960.5 & 2700.1 & 4954.8 & 11105 & 10 & 15 & 3600.0 & 4400.0 & 6330.0 & 8330.0\\
    05089 & 10 & 15 & 2593.9 & 3170.3 & 6267.6 & 7708.7 & 11108 & 10 & 15 & 2604.6 & 3183.4 & 8001.9 & 9448.9\\
    05135 & 10 & 15 & 5201.7 & 6357.6 & 1992.2 & 4882.0 & 11442 & 10 & 15 & 4201.7 & 5602.2 & 6559.9 & 8894.2\\
    05152 & 10 & 15 & 5941.6 & 7262.0 & 2478.7 & 5779.6 & 11522 & 10 & 15 & 4544.0 & 5553.8 & 7085.9 & 9610.3\\
    05239 & 10 & 15 & 2949.7 & 3605.1 & 4044.1 & 6502.2 & 11683 & 10 & 15 & 4530.3 & 5537.0 & 5684.8 & 8201.6\\
    05949 & 10 & 15 & 2935.1 & 3587.4 & 5223.3 & 6853.9 & 11730 &  8 & 10 & 4715.8 & 6287.7 & 7017.5 & 9637.4\\
    06498 & 10 & 15 & 2518.7 & 3078.4 & 5812.7 & 7212.0 & 11833 & 10 & 20 & 2971.9 & 3632.3 & 5065.1 & 6716.2\\
    06909 &  6 &  8 & 3798.2 & 4642.3 & 5445.9 & 7556.0 & 11966 & 10 & 15 & 4537.6 & 6806.5 & 5910.4 & 8746.4\\
    06981 & 10 & 15 & 3834.4 & 4686.5 & 4204.9 & 6335.1 & 11989 & 10 & 15 & 3711.3 & 4536.0 & 7184.4 & 9246.3\\
    06992 & 10 & 15 & 3952.6 & 5270.1 & 3434.1 & 5630.0 & 11990 & 10 & 15 & 1721.4 & 2104.0 & 5835.8 & 7270.3\\
    07051 & 10 & 15 & 2035.9 & 2488.4 & 6224.9 & 7356.0 & 12333 & 10 & 15 & 2944.6 & 3599.0 & 5382.4 & 7563.6\\
    07079 & 10 & 15 & 6042.1 & 7384.8 & 6435.9 & 9792.6 & 12343 & 10 & 15 & 2595.1 & 3892.6 & 3979.3 & 5601.2\\
    07275 & 10 & 15 & 4800.0 & 9600.0 & 4119.1 & 8119.1 & 12466 & 10 & 15 & 3275.2 & 4912.7 & 3155.0 & 5543.2\\
    07566 & 10 & 15 & 3167.7 & 3871.6 & 4704.4 & 6464.2 & 12664 & 10 & 15 & 4131.6 & 5049.8 & 5331.6 & 7627.0\\
    07638 & 10 & 15 & 2050.7 & 2506.4 & 5955.4 & 7094.7 & 12786 & 10 & 15 & 4391.8 & 5367.8 & 6876.5 & 9316.4\\
    07640 & 10 & 15 & 1133.7 & 1385.7 & 7960.1 & 8589.9 & 12819 & 10 & 15 & 2680.4 & 4467.3 & 6036.1 & 8642.0\\
    07659 & 10 & 15 & 4302.9 & 5737.2 & 7058.8 & 9847.7 & 13159 & 10 & 15 & 3145.7 & 3844.8 & 3602.5 & 5350.1\\
    07943 & 10 & 15 & 3437.0 & 4200.8 & 5534.8 & 7444.2 & 13260 & 10 & 15 & 2908.8 & 3555.2 & 6839.4 & 8455.5\\
    07995 & 10 & 15 & 5007.2 & 7232.7 & 5026.8 & 7808.6 & 13454 & 10 & 15 & 961.4 & 1175.0 & 7641.2 & 8175.3\\
    08185 & 10 & 15 & 4199.1 & 5132.2 & 3974.5 & 6307.3 & 13477 & 10 & 15 & 4251.4 & 5196.1 & 4491.5 & 6853.4\\
    08215 & 10 & 15 & 5977.7 & 10461.0 & 3753.0 & 7489.1 & 13534 & 10 & 15 & 3277.2 & 4005.5 & 6515.3 & 8336.0\\
    08442 & 10 & 15 & 3056.8 & 3736.0 & 5005.3 & 6703.5 & 13896 & 10 & 15 & 2675.8 & 3270.4 & 7895.0 & 9381.6\\
    08697 & 10 & 15 & 2061.7 & 2519.8 & 6684.2 & 7829.5 & 14101 & 10 & 15 & 3691.4 & 4511.7 & 6782.7 & 8833.5\\
    08752 & 10 & 15 & 3515.0 & 4296.1 & 5113.8 & 7066.6 & 14145 & 10 & 15 & 3677.7 & 4495.0 & 6302.5 & 8345.7\\
    08830 & 10 & 15 & 4295.0 & 5249.5 & 4148.5 & 6534.6 & 14526 & 10 & 15 & 2621.2 & 3203.7 & 5453.9 & 6910.1\\
    09104 & 10 & 15 & 4944.5 & 6043.3 & 4733.3 & 7480.2 & 14784 &  8 & 15 & 2353.2 & 3529.8 & 5981.7 & 8187.8\\
    09276 & 10 & 15 & 4661.6 & 5697.5 & 5394.1 & 7983.8 & 14786 & 10 & 15 & 2078.6 & 2540.5 & 4644.0 & 5798.8\\
    09577 & 10 & 15 & 1921.2 & 2348.2 & 5804.4 & 6871.7 & 14815 & 10 & 15 & 2441.2 & 3255.0 & 5550.9 & 6907.1\\
    09635 & 10 & 15 & 3309.2 & 4780.0 & 6280.1 & 8118.6 & 14830 & 10 & 15 & 3918.7 & 5225.0 & 5831.3 & 8008.3\\
    09683 & 10 & 15 & 4222.5 & 6099.2 & 5900.0 & 8245.9 & 14891 & 10 & 15 & 5695.9 & 8227.4 & 6526.9 & 9691.2\\
    09698 & 10 & 15 & 3985.7 & 6293.2 & 5848.3 & 7200.0 & 14905 & 10 & 15 & 3373.1 & 4122.7 & 6855.6 & 8729.6\\
    09778 & 10 & 15 & 2734.2 & 3341.7 & 7648.8 & 9420.9 & 15388 & 10 & 15 & 2626.9 & 3210.7 & 7568.6 & 9028.0\\
    09781 & 15 & 20 & 2938.9 & 4041.0 & 5489.7 & 7326.5 & 15394 & 10 & 15 & 3159.6 & 3861.7 & 6514.7 & 8855.1\\
    09789 & 10 & 15 & 3337.6 & 4079.3 & 7241.4 & 9713.7 & 15784 & 10 & 15 & 2896.6 & 4344.8 & 6714.5 & 8826.6\\
    10047 & 10 & 15 & 3530.5 & 5099.7 & 3708.5 & 5669.9 & 15841 & 10 & 15 & 3003.5 & 3671.0 & 2900.0 & 5400.0\\
    10142 & 10 & 15 & 3278.4 & 4507.8 & 4891.7 & 6940.7 & 20376 &  5 & 10 & 1483.6 & 1813.3 & 7241.2 & 8065.4\\
    10158 & 10 & 15 & 3548.5 & 5322.7 & 4572.6 & 6790.4 & 20627 & 10 & 15 & 1069.1 & 1306.7 & 7195.4 & 7888.4\\
    10210 & 10 & 15 & 3144.5 & 3843.3 & 6016.1 & 7763.0 & 31312 & 10 & 15 & 1860.2 & 2273.6 & 7019.7 & 8053.2\\
    10356 & 10 & 15 & 3788.6 & 5682.9 & 4583.8 & 7740.9 & --    & -- & -- & --     & --     & --     & --\\
    \enddata
\end{deluxetable*}

A critical step in the application of Bayesian statistics is the choice of prior distributions. 
For the three parameters of the parabolic component in Eq.~\ref{eq:parabola}, $c_0$, $c_1$, and $c_2$, we apply uniform priors within the ranges of
[-1; 1], [$-10^{-4}$; 10$^{-4}$], and [$-10^{-8}$; $10^{-8}$], respectively. The projected semi-major axis $a_1 \sin i$ is also constrained
with a uniform prior to [0;10] (AU) for most stars, with the only exception being 20376, where this was modified to [0;5] (au) to avoid
degenerate posterior samples with really large $a_1 \sin i$ values.

In order to determine appropriate ranges for the priors on the orbital period and the time of periastron passage, we make use of their
values determined during the first iteration of the $O-C$ method: for the former, we adopt a default uniform prior in the range
[$0.9\times P_{\mathrm{orb},1}$; $1.1\times P_{\mathrm{orb},1}$]; for the latter, in
[$T_{0,1} - 0.25 \times P_{\mathrm{orb},1}$; $T_{0,1} + 0.25 \times P_{\mathrm{orb},1}$], where $P_{\mathrm{orb},1}$ and $T_{0,1}$
are the estimates for the quantities from the first iteration of the $O-C$ method. It should be noted that $T_0$ is a periodic (circular)
quantity with a period of $P_{\mathrm{orb}}$, hence the choice of these (default) constraints.
We adopted flat priors of [-1;1] on the transformed quantities $\sqrt{e} \sin \omega$ and $\sqrt{e} \cos \omega$,
resulting in an effective flat prior on $e$ in the range of [0;1] (after converting them through Eq.~\ref{eq:econv}).

There are two additional implicit ``priors'' utilized during the analysis, that is, they are necessary to reproduce the MCMC
fits presented in Table~\ref{tab:binprop} and Figure~\ref{fig:ocs}: the orders of the Fourier series used during the first and
second iterations of the $O-C$ method employed here. Therefore, we list these parameters, along with the uniform prior ranges on
$P_\mathrm{orb}$ and $T_{0}$, in Table~\ref{tab:priors}.

\bibliography{rrl_bin_1}{}
\bibliographystyle{aasjournal}

\end{document}